\documentstyle[12pt,aaspp4]{article}

\begin{document}

\title{Global Relationships Among the Physical Properties of Stellar Systems}
\author{David Burstein}
\affil{Department of Physics and Astronomy,\\Box 871054, Arizona State 
University\\ Tempe, AZ 85287-1504}

\author{Ralf Bender}
\affil{Universit\"{a}ts-Sternwarte M\"{u}nchen\\Scheinerstrasse 1\\
D-81679 M\"{u}nchen, Germany}

\author{S. M. Faber and R. Nolthenius}
\affil{UCO/Lick Observatory \\University of California\\
Santa Cruz, CA 95064  U.S.A.}

\begin{abstract}

The $\kappa$-space three-dimensional parameter system was originally defined
to examine the physical properties of dynamically hot elliptical galaxies and
bulges (DHGs). The axes of $\kappa$-space are proportional to the logarithm of
galaxy mass, mass-to-light ratio, and a third quantity that is mainly surface
brightness. In this paper we define self-consistent $\kappa$ parameters for
disk galaxies, galaxy groups and clusters, and globular clusters and use them
to project an integrated view of the major classes of self-gravitating,
equilibrium stellar systems in the universe. Each type of stellar system is
found to populate its own fundamental plane in $\kappa$-space. At least six
different planes are found: 1) the original fundamental plane for DHGs; 2) a
nearly--parallel plane slightly offset for Sa-Sc spirals; 3) a plane with
different tilt but similar zero point for Scd-Irr galaxies; 4) a plane
parallel to the DHG plane but offset by a factor of 10 in mass-to-light ratio
for rich galaxy clusters; 5) a plane for galaxy groups that bridges
the gap between rich clusters and galaxies; and 6) a plane for Galactic
globular clusters.  We propose the term ``cosmic metaplane'' to describe this
ensemble of interrelated and interconnected fundamental planes.

The projection $\kappa_1$--$\kappa_3$ ($M/L$ vs. $M$) views all planes
essentially edge-on. Planes share the common characteristic that $M/L$ is
either constant or increasing with mass. The $\kappa_1$--$\kappa_2$ projection
views all of these planes close to face-on, while $\kappa_2$--$\kappa_3$ shows
variable slopes for different groups owing to the slightly different tilts of
the individual planes.  The Tully-Fisher relation is the correct compromise
projection to view the spiral-irregular planes nearly edge on, analogous to
the $D_n$-$\sigma$ relation for DHGs.  No stellar system yet violates the rule
first found from the study of DHGs, namely, $\kappa_1 + \kappa_2 <$ constant,
here chosen to be 8.  In physical terms, this says that the maximum global 
luminosity density of stellar systems varies as $M^{-4/3}$.  Galaxies march 
away from this ``zone of exclusion" (ZOE) in $\kappa_1$--$\kappa_2$ as a 
function of Hubble type: DHGs are closest, with Sm--Irr's being furthest away.

The distribution of systems in $\kappa$-space is generally consistent with
predictions of galaxy formation via hierarchical clustering and merging. The
cosmic metaplane is simply the cosmic virial plane common to all
self-gravitating stellar systems, tilted and displaced in mass-to-light ratio
for various types of systems due to differences in stellar population and
amount of baryonic dissipation. Hierarchical clustering from an $n = -1.8$
power-law density fluctuation spectrum (plus dissipation) comes close to
reproducing the slope of the ZOE, and the progressive displacement of Hubble
types from this line is consistent with the formation of early-type galaxies
from higher $n$-$\sigma$ fluctuations than late Hubble types.

The $M/L$ values for galaxy groups containing only a few, mostly spiral
galaxies, vary the strongest with $M$.  Moreover, it is these groups that
bridge the gap between the two planes defined by the brightest galaxies and 
the lowest mass rich clusters, giving the cosmic metaplane its striking 
appearance.  Why this is so is but one of four key questions raised by 
our study.  The second question is why the slopes of {\it individual} Hubble 
types in the $\kappa_1$--$\kappa_2$ lie plane parallel the ZOE.  At face 
value, this appears to suggest {\it less} dissipation of massive galaxies 
within their dark halos compared to lower-mass galaxies of the same Hubble 
type.  The third is why we find isotropic stellar systems only within an
effective mass range of $\rm 10^{9.5-11.75} \, M_\odot$.  This would seem
to imply that dissipation only results in galaxy components flattened by
rotation in a limited mass range.  The fourth question, perhaps the most
basic of all, is how does $M/L$ vary so smoothly with $M$ among all stellar
systems so as to give the individual tilts of the various fundamental planes,
yet preserve the overall appearance of a metaplane?  The answer to this
last question must await a more thorough knowledge of how galaxies relate
to many parameters, including: their environment, structure, angular momentum
acquisition, density, dark matter concentration, the physics of star formation
in general, and the formation of the initial mass function in particular.

The present investigation is limited by existing data to the B passband and is
strongly magnitude-limited, not volume-limited.  Rare or hard-to-discover
galaxy types, such as H II galaxies, starburst galaxies and
low-surface-brightness galaxies, are missing or are under-represented, and use
of the B band over-emphasizes stellar population differences.  A
volume-limited $\kappa$-space survey based on K-band photometry and complete
to low surface brightness and faint magnitudes is highly desirable but
requires data yet to be obtained.

\end{abstract}


\section{Introduction}

It is now well established that orbital velocity increases strongly with
galaxy luminosity in giant galaxies of all Hubble types.  This relationship
is embodied in the Tully--Fisher (Tully \& Fisher 1977) relationship for
spiral and irregular galaxies, and the fundamental plane (Dressler et al.
1987; Djorgovski \& Davis 1987) for elliptical galaxies and spiral bulges
(Dressler 1988; Bender, Burstein \& Faber 1992; hereafter $\rm B^2F1$).

In $\rm B^2F1$ we developed a new way of viewing the structural properties of
dynamically hot galaxies (hereafter DHGs) by making an orthogonal rotation of
the global parameter space defined by velocity dispersion ($\sigma$), effective
radius ($r_e$) and mean effective surface brightness within $r_e$ ($I_e$).
This rotation leads to a new coordinate system that we named ``$\kappa$-space''
and which provides face-on and edge-on views of the DHG fundamental plane. The
axes of $\kappa$-space are logarithmically related to galaxy mass,
mass-to-light ratio, and a third quantity that depends primarily on surface
brightness (the $\kappa$ parameters are defined in \S~\ref{dat-2}).

$\kappa$-space is a useful diagnostic tool for visualizing the physical
properties of DHGs. It is therefore natural to ask whether an analogous plot
for other Hubble types, and even galaxy groups and clusters, could be
similarly informative.

Earlier discussions of the relationships among the physical properties of
stellar systems concentrated on scaling relationships between just two
parameters at a time. Examples are the Tully--Fisher (TF) relation for spirals
and the Faber--Jackson (1976) relation for ellipticals: $L \propto V^4$.  We
now understand that, for elliptical galaxies, $L \propto V^4$ is just one
projection of what is inherently a two-dimensional family, the fundamental
plane (FP). It is natural to conjecture that the TF relation is a similar
projection for spiral and irregular galaxies, but this has not yet been shown
explicitly.  If there is a FP for spirals and irregulars, what is its relation
to the DHG FP, and how do the various Hubble types distribute themselves
within it?

The various scaling relations for stellar systems provide a test for theories
of structure formation in the the universe. Faber (1982a,b) and Blumenthal et
al. (1984; hereafter BFPR), among others, examined the projected distributions
of galaxies and clusters onto the (density, velocity) and (mass,velocity)
planes and compared them to the predictions of hierarchical clustering theory.
The distributions of rich galaxy clusters (Schaeffer et al. 1993) and
globular clusters (Djorgovski 1995) were examined in standard FP space.  These
latter were found to form planes nearly parallel to the DHG FP.

Despite these individual investigations, there still has not been a unified
survey of all types of self-gravitating, equilibrium stellar systems in the
three-dimensional space of their fundamental structural parameters.  In this
paper we assemble the necessary database to provide such a survey.  A
preliminary version of this work was presented in Burstein et al. (1995;
hereafter $\rm B^2FN$).  Here we describe how our homogeneous and
self-consistent data set has been assembled and how the structures of the major
groups of self-gravitating stellar systems fit together in $\kappa$-space. 
We discuss how their overall distribution can be interpreted in terms of
hierarchical clustering theory.  A parallel goal of this effort is to make
available a benchmark data set against which the success of models for
structure formation in the universe can be judged.  Finally, our catalog of
local structural parameters is a foil for similar structures at high redshift,
for the purpose of detecting and quantifying the evolution of structure in the
universe using the lookback effect.

The database contains the most familiar self-gravitating, virialized or
nearly-virialized systems in the nearby universe, including galaxies, globular
clusters, groups of galaxies, and large clusters of galaxies.  However, rare
types of galaxies are missing or underrepresented, e.g. starburst H II
galaxies (e.g., Salzer et al. 1989; Gallego et al. 1996), compact, narrow
emission line galaxies (Koo et al. 1994) and very low surface brightness
galaxies (e.g., McGaugh 1996; Dalcanton 1997).  Thus, the true volume
populated by galaxies in $\kappa$-space is likely to be more extended to both
lower and higher surface brightnesses than shown here.  Our galaxy sample is
also approximately magnitude-limited, so that the number density versus
position in $\kappa$-space is not an accurate picture of the local volume
densities of galaxies.  Even with these shortcomings, the present data are a
useful first overview of the properties of common structures in the nearby
universe.

Definitions of $\kappa$-space parameters are reviewed in \S~\ref{dat} and
explicitly equated to typical galaxy parameters such as mass, radius and 
luminosity in
Appendix~\ref{app-1}.  In \S~\ref{kap1} we develop a more appropriate way
(than used in $\rm B^2FN$) to intercompare the properties of galaxies of
different morphological types, as well as to homogenize the properties of
galaxy groups, galaxy clusters and globular clusters.  \S~\ref{kapgal} is a
picture gallery that shows, in various projections, how different types of
galaxies are distributed within $\kappa$-space.  The $\kappa$-space
distributions of galaxy groups, galaxy clusters and globular clusters are
compared to those of galaxies in \S~\ref{kapgp}, and tested with N-body models
in Appendix~\ref{app-2}. In \S~\ref{kaphcm} we discuss how these data accord
with the current picture of hierarchical structure formation in the universe.
Our results are summarized in \S~\ref{kapconcl}.

\section{Data}
\label{dat}

\subsection{Data Needed for $\kappa$ Parameters}
\label{dat-1}

Following $\rm B^2F1$, we choose to define the photometric parameters of
stellar systems using the B passband.  This choice is dictated by the large
database of available B photometry. It would clearly be valuable to see how
these properties relate in another passband, particularly one less sensitive
to stellar population age such as near-infrared K.  Unfortunately, the data do
not yet exist to do so, nor are they likely to appear soon in the quantity
currently available for B-band data.

The global structure of self-gravitating, equilibrium stellar systems can be
characterized by three parameters: size, surface brightness, and internal
orbital velocity. We use the photometric data to measure the radius containing
half the luminosity (the effective radius, $r_e$) and the mean surface
brightness within this effective radius ($I_e$).  As a measure of orbital
velocity we use the {\it one-dimensional central} velocity dispersion
($\sigma_c$). This is arbitrary, as arguments could be made to use the mean
velocity dispersion within the effective radius for galaxy clusters, say, or
mean rotation speed for spiral and irregular galaxies. Our choice of central
velocity dispersion follows the convention of $\rm B^2F1$, but we will need to
generate this parameter from other types of orbital velocity measurements (see
below). This section presents the basic sources of initial data. In
\S~\ref{kap1} we discuss how the data are placed onto a common system.

\subsection{Definition of $\kappa$-Parameters}
\label{dat-2}

The $\kappa$-space parameters are defined as follows:

\begin{equation}
\kappa_1 \quad \equiv \quad (\log {\sigma_c}^2 \, + \, \log r_e)/\sqrt{2},
\end{equation}

\begin{equation}
\kappa_2 \quad \equiv \quad (\log \sigma_c^2 \, + \,2 \log I_e \, - \,
                                               \log r_e)/\sqrt{6}
\end{equation}

\centerline{and}

\begin{equation}
\kappa_3 \quad \equiv \quad (\log \sigma_c^2 \, - \, \log I_e \, - \,
                                               \log r_e)/\sqrt{3}.
\end{equation}

\noindent
The quantity $\sigma_c$ is the central velocity dispersion in km s$^{-1}$.
As in $\rm B^2F1$, $r_e$ is measured in kpc, while $I_e$ is measured in
units of B-band solar luminosities pc$^{-2}$, computed as $10^{-0.4 \cdot
(SB_e-27.0)}$ where $SB_e$ is the mean B-band surface brightness in B mag
arcsec$^{-2}$ within $r_e$.  

In many cases it useful to have convenient transformations between the
$\kappa$ parameters and such typical parameters for stellar systems such as
mass, luminosity, size, mass-to-light ratio, etc.  In the present paper
we use such transformations in many ways, in order to interpret the properties
of stellar sytems.  As such, in Appendix~\ref{app-1} we give the equations
that transform the $\kappa$ parameters to these common parameters, including
all the other parameters we employ in this paper.

\subsection{Distances}
\label{dat-3}

Distances for galaxies, galaxy groups and galaxy clusters are determined using
Local Group radial velocities and an assumed smooth Hubble flow with $H_0 =
50$ km sec$^{-1}$ Mpc$^{-1}$ (cf. $\rm B^2F1$).  In adopting this prescription
we are purposefully ignoring real peculiar motions, as these motions would
currently have to be derived from the very physical properties of the stellar
systems we wish to study.  This approach is adequate as our goal is to discuss
the physical relationships among stellar systems in broad-brush strokes. 
Distances for ellipticals, bulges and dwarf ellipticals are based on their
group or individual radial velocities as tabulated by Faber et al. (1989) and
$\rm B^2F1$.  Most spiral and irregular galaxy distances are based on
individual Local Group radial velocities as given in de Vaucouleurs et al.
(1991, hereafter RC3).

There are three exceptions to the use of a smooth Hubble flow and Local Group
radial velocities.  Fifteen spirals and all DHGs whose radial velocities and
structural properties place them in the virialized parts of Virgo or Coma are
given the distances of their respective clusters.  The Virgo cluster distance
is assumed to be 20.7 Mpc (based on an assumed Local Group velocity for the
cluster of 1035 km sec$^{-1}$).  The Coma cluster distance is as in $\rm
B^2F1$, 137.9 Mpc, based on a Local Group velocity of 6,895 km sec$^{-1}$. 
Local Group DHGs are given distances as in $\rm B^2F1$.  Local Group spiral
and irregular galaxies are put either at the generally--accepted distance of
M31 (0.7 Mpc) or of M33 (1 Mpc). Globular cluster distances in our Galaxy are
determined from the V mag of the horizontal branch, as listed by Peterson 
(1993).

\subsection{Dynamically Hot Galaxies}
\label{dat-4}

The $\kappa$ parameters for DHGs with measures of internal orbital anisotropy
are given in $\rm B^2F1$.  The reader is referred to that paper for details.
The present catalog contains 292 additional elliptical galaxies from Faber et
al. (1989) lacking anisotropy measures and not included in $\rm B^2F1$.

\subsection{Spiral and Irregular Galaxies}
\label{dat-5}

Observed $\kappa$ parameters for spiral and irregular galaxies are derived
from data in the RC3. A search was made of the computer-readable copy of the
RC3 (kindly supplied by H.G. Corwin, Jr.) for galaxies with listed values of
radial velocity, B-band values of $SB_e$, angular radius $\theta_e$ and
observed velocity widths $\Delta V$ ($W_{20}$ in RC3 terminology).  We accept 
the RC3 definitions of $\Delta V$, inclination and effective radius and define
rotation velocity as $V_{rot} = \Delta V/2 \sin i$ (where $\sin i =
\arccos[\sqrt{(b/a)^2 - 0.04)/0.96}]$ and $a/b$ is the axial ratio taken from
the RC3). Known errors in determining $V_{rot}$ are at the 10\% level (RC3),
comparable to the errors of velocity dispersions (Davies et al. 1987).

From this first cut, we exclude all galaxies labeled as ``peculiar,'' those
with inclinations more face-on than $45^{\circ}$, and those with Galactic
extinction $A_B > 0.75$ mag, as predicted by the method of Burstein \& Heiles
(1978, 1982).  We also exclude three Sa galaxies (UGC 1344, NGC 4424 and NGC
7232) for which the tabulated values of $V_{rot}$ are apparently in error 
(too low).  This leaves 511 galaxies of Hubble types Sa through Im.

To derive mean effective surface brightness and radius we have corrected the
RC3 values of circular effective aperture size ($A_e$) and mean surface
brightness within $A_e$ (termed ${m'}_e$ in the RC3) using the
axial-ratio-based statistical relationships defined in the RC3.  In addition,
the surface brightnesses of all spiral galaxies have been corrected for
Galactic extinction and for internal extinction by subtracting 1.5 log ($a/b$)
mag. This is an average correction taken from the RC3 analysis.  The
inclination-based corrections to translate $A_e$ and ${m_e}'$ to $r_e$ and
$I_e$ are typically small (0.04 dex and 0.08 dex respectively for log ($a/b$)
= 0.4). Internal extinction corrections are larger ($\pm 0.3$ mag) for a range
of inclinations corresponding to log ($a/b$) = 0.4 to 0.8.  The internal
corrections for extinction correct the galaxy to face-on.

The effective photometric parameters for spiral galaxies from the RC3 are
affected by errors in axial ratio, $A_e$ and ${m_e}'$, which combine to
produce errors of order $>$0.1 dex in $r_e$ and $SB_e$.  Since such errors
dwarf either K-corrections ($<$0.04 dex) or cosmological corrections, ($<$0.03
dex to $SB_e$), we choose not to make these further corrections for the spiral
galaxy data.  As we will see, errors of 0.1 dex have no effect on our
conclusions.

In addition to the RC3 data, $\kappa$ parameters are derived for two large,
low surface brightness spiral galaxies identified by Impey \& Bothun (1989;
Malin 1) and by Bothun et al. (1990; which we call here Malin 2). Rotation
velocity, effective radii and effective surface brightnesses are derived from
data in these papers, with distances adjusted to our assumed value of $H_0 =
50$ km sec$^{-1}$ Mpc$^{-1}$.  Values of $r_e$ and $SB_e$ for these two
galaxies are probably no more accurate than 0.15 dex, and the value for
rotation velocity no better than 15\%. As above, these uncertainties have no
effect on our conclusions.

\subsection{Galaxy Groups}
\label{dat-6}

The only genuinely new data in this paper are structural parameters for galaxy
groups (as opposed to clusters, see below). Appendix~\ref{app-2} shows 
selecting groups in real-space vs. redshift space changes group parameters by a 
systematic but very small amount, compared to the overall spread in physical
properties among groups, and does not affect our conclusions.

The galaxy groups here come from the {\it Nearby Groups Catalog} of Nolthenius
(1993; hereafter N93).  The N93 groups were identified from the $m_{lim}=14.5$
CfA1 catalog of Davis et al. (1982), and those used here have at least 3
members.  Following Geller \& Huchra (1983) and Nolthenius \& White (1987),
the N93 groups are identified by a percolation algorithm linking galaxies on
the sky and in redshift so that the number overdensity threshold should be
constant with redshift.  In N93 the sky separation linking criterion ($D_0$)
is chosen to be less than 0.36 of the mean intergalaxy separation, which
corresponds to a limiting number overdensity of $\ge$ 20. The redshift link
scales with depth according to sample completeness and is 350 km sec$^{-1}$ at
a distance of 5000 km sec$^{-1}$ (corresponding to a redshift link parameter
$V_5 = 350$; see N93).

For computing group surface brightnesses and $M/L$, we have used
fully-corrected $B_T(0)$ magnitudes for the individual galaxies on the RC3
system.  These are available for 96\% of the galaxies.  This ensures that the
photometric $\kappa$ parameters  for the individual galaxies and groups are on
the same system. Only 45 of the 1095 galaxies in 170 groups (excluding the
Abell 194 and Coma clusters, which are also in the Schaeffer et al. (1993)
cluster sample) do {\it not} have quoted RC3 values of $B_T(0)$.  For these
45 galaxies alone we use the $B_T$ values given by N93.

The half light radius $r_e$ for a group is found by taking a
luminosity-weighted centroid of the visible galaxies, and the radius enclosing
half the visible light is interpolated.   We considered unweighted centroids
and found that the extent and tightness of groups in $\kappa$-space was
virtually identical.  We prefer $L$-weighted centroids as these more closely
parallel the method used to define the photocenters of galaxies.   To be
further consistent with the parameters for galaxies, other sources of baryonic
flux (e.g., X-ray emitting hot gas) are not considered. The values of $r_e$
and $SB_e$ for galaxy groups are probably accurate at the 0.3 dex level.

Velocity dispersions for the $n$-visible galaxies in a group with redshifts 
$v_i$ are found from

\begin{equation}
\sigma^{2} = {1\over{(n-1)}}\sum_{i}^n {(v_i - \overline v)}^2.
\end{equation}

\noindent Velocity dispersions determined in this manner refer essentially
to the effective radius of the group, $\sigma_e$.  In contrast, the velocity
dispersions for DHGs are measured at their centers, $\sigma_c$.  We take 
account of this difference in \S~\ref{kap1-2}.

In addition to the observed $\kappa$ parameters and related data for groups,
we also calculate the fraction of E$+$S0 galaxies. This fraction can be taken
as the analog of Hubble type for individual galaxies.   An E/S0 fraction of
1/3 is taken as the dividing line between E-rich ($>$ 1/3 E/S0's) and E-poor 
($\le$ 1/3 E/S0's) groups.

\subsection{Galaxy Clusters}
\label{dat-7}

Schaeffer et al. (1993) have shown that the physical properties of 16 rich
clusters of galaxies define a fundamental plane similar in slope but offset in
zero point from that defined by elliptical galaxies.  A. Cappi kindly supplied
us with a computer-readable list of the data used by Schaeffer et al. These
velocity dispersions refer approximately to $r_e$, so we correct them to
$\sigma_c$ as for groups (see \S~\ref{kap1-2}). $\kappa$ parameters are
calculated after revising the Schaeffer et al. distance scale to $H_0$ = 50.
Since the Schaeffer et al. clusters are among the richest nearby large clusters,
we assume that all are E-rich by our group-based criterion.

\subsection{Globular Clusters}
\label{dat-8}

$\kappa$ parameters for 51 globular clusters are based on tables given in the
appendices to ``Structure and Dynamics of Globular Clusters'' (Djorgovski \&
Meylan 1993).  Half light radii are taken from Trager, Djorgovski
\& King (1993), $\sigma_e$ from Pryor \& Meylan (1993), and distances,
$E(B-V)$, total $V$ magnitude, and $B-V$ colors from Peterson (1993). The
effective radius is the half light radius given by Trager et al., and the
effective $V$ magnitude is 0.75 mag ($\times 2$) fainter than the total
$V$ magnitude listed by Peterson. B-band surface brightness (to derive
$I_e$) is obtained by correcting both $B-V$ and $V$ for reddening and
extinction using $A_V = 3 E(B-V)$. Most of the globular clusters have
reliable measurements of half light radii and total $V$ mag; for several, one
of these quantities is more uncertain than the other.  Clusters are excluded
that have poor observational data for both quantities, as these are seen to
scatter more within $\kappa$-space.

\section{Homogeneous $\kappa$-space Parameters for Different 
Stellar Systems}
\label{kap1}

\subsection{Effective Radius and Effective Surface Brightness}
\label{kap1-1}

The definition of $r_e$ as the radius encompassing half the light is standard
for all galaxy types and globular clusters and is straightforwardly extended
to galaxy groups and clusters (see above). Ideally, we would wish that our
values of $r_e$ and $I_e$ referred to the {\it total} baryonic mass
distribution, not just that of the stars, but by using B-band light we are
ignoring all non-stellar baryons. For dwarf irregular galaxies, B-band $r_e$
values are likely to underestimate the baryonic half-mass radius since many
such galaxies have extended H~I gas distributions (e.g. DDO~154, Krumm \&
Burstein 1984). $I_e$ likewise underestimates the baryonic surface density by
ignoring H~I and H$_2$. Values of $r_e$ and $I_e$ for groups and clusters
ignore the distribution of hot X-ray emitting gas known to be a major baryonic
component of clusters (e.g., Fabian 1994) and more recently found in smaller
E-rich galaxy groups (e.g., Mulchaey et al. 1996).  As the X-ray gas appears
to be more extended than the galaxies (Mulchaey et al. 1996), this could lead
to underestimation of baryonic values of $r_e$ for these stellar systems by
20--50\%, and underestimation of baryonic surface density by up to a 
factor of 3.

Even for stars, the B band is susceptible to perturbing effects. B-band
parameters are reasonably consistent for DHGs, globular clusters, and
E-dominated groups and clusters, but the light of spiral and irregular
galaxies (and spiral-dominated groups/clusters) is strongly perturbed by dust
and young stars. While we might attempt to correct $I_e$ between spirals and
DHGs for stellar population-related mass-to-light differences, we choose not
to do so for two reasons.  First, the range of stellar populations among
spirals of similar Hubble type is nearly as large as that between spiral
galaxies and elliptical galaxies (Burstein 1982).  Second, models of stellar
populations (Worthey 1994) indicate that such corrections typically change
$M/L$ in the B passband by $\sim \times 2$, leading to $\kappa_3$ errors of
less than 0.3 dex.

The presence of the above systematic trends and omissions is clearly evident
in the data, most obviously in values of $M/L$ for different types of objects.
However, present knowledge does not permit us to account for these effects in
any reliable way.  To compile the survey, we prefer to stay firmly rooted in a
well defined and internally consistent database (B-band photometry) and leave
such corrections to the future.

\subsection{Characteristic Internal Velocity}
\label{kap1-2}

The most difficult parameter to define homogeneously for different types of
stellar systems is characteristic internal velocity.  In keeping with our
philosophy, we would like to define a velocity that is simply related to a
system's mass. This $\sigma_{ideal}$ would combine with $r_e$ to yield a mass
that is consistently measured for all stellar systems.  How to define such a
$\sigma_{ideal}$ is not clear, but we can operationally define transformations
among common velocity systems to bring them reasonably close onto the same
system.

.For the present sample, there are three kinds of characteristic velocities: i)
$\sigma_c$, the central velocity dispersion, used for giant E galaxies and the
bulges of spiral galaxies ($\rm B^2F1$); ii) $\sigma_e$, the velocity
dispersion near within $r_e$, used for globular clusters, galaxy groups and 
galaxy clusters; and iii) $V_{rot}$, the maximum circular rotation velocity 
for disk galaxies. $\sigma_c$ and $\sigma_e$ are line-of-sight, 1-D measures of
dispersion, which is correlated with the local profile slope as well as mass,
whereas $V_{rot}$ relates mainly to enclosed mass alone.  The goal is to
identify relationships among these three parameters that will, on average,
permit us to transform one measurement into another to an accuracy of $\sim
0.1$ dex. Two such transformations are needed: $\log \sigma_c = \log \sigma_e
+ K_1$ and $\log \sigma_c = \log V_{rot} + K_2$, where the task is to estimate
the values of $K_1$ and $K_2$.

The relationship between $\sigma_e$ and $\sigma_c$ is known for King (1966)
and Jaffe (1983) models (Appendix A of $\rm B^2F1$; Chapter 4 in
Binney \& Tremaine 1987).  A reasonable compromise ($\pm 10$\%) is
$\sigma_e/\sigma_c = 0.84$, yielding $K_1 = +0.076$ dex. We assume that
globular clusters and galaxy groups and clusters obey King-Jaffe models.  The
additive corrections to be applied to the $\kappa$ parameters from this
correction are small: $\Delta_1(\kappa_1) = 0.11$ dex, $\Delta_1(\kappa_2) =
0.06$ dex, and $\Delta_1(\kappa_3) = 0.09$ dex.

Should we also consider making a correction for anisotropy in the velocity
dispersion?  Before doing so, one should recognize that there are really two
kinds of anisotropy.  One is {\it x--y--z} anisotropy, which leads to a range
of flattenings and triaxialities as, for example, in luminous E galaxies ($\rm
B^2F1$).  The other is determined by the ratio of radial-to-tangential orbital
motions. These two anisotropies can coexist in the same object.  Fortunately,
it seems that {\it x--y--z} anisotropy is benign in $\kappa$-space, as shown
by the fact that apparent galaxy ellipticity plays a negligible role in the
scatter of DHGs relative to the fundamental plane (cf. Saglia, Bender \&
Dressler 1992; J{\o}rgensen, Franx \& Kjaergaard 1996). Little information is
available on radial-to-tangential anisotropy, but hot systems that formed via
violent relaxation should not be strongly radially anisotropic at the 10\%
level of accuracy needed here. Hence we make no corrections for either type of
anisotropy to the velocity dispersions of DHGs.

For spiral and irregular galaxies with exponential Freeman (1970) disks and
central surface brightnesses equal to 21.65 B mag arcsec$^{-2}$, the effective
radius occurs at 23.47 B mag arcsec$^{-2}$.  Among galaxies with central
surface brightnesses within 1~mag arcsec$^{-2}$ of this value, most objects
have reached their observed maximum rotation velocity by $r_e$ or are close to
it (Rubin et al. 1985), so it is reasonable to assume that $V_{rot}$ is
measured at $r_e$.  For disks embedded in isotropic isothermal halos, we
expect $V_{rot}/\sigma_c = \sqrt{2}$ (Binney \& Tremaine, 1987, Eq. 4--55),
which agrees with observed values (Whitmore \& Kirshner 1981).  We therefore
set $K_2 = -0.151$ dex and find $\Delta_2(\kappa_1) = -0.21$ dex,
$\Delta_2(\kappa_2) = -0.12$ dex, and $\Delta_2(\kappa_3) = -0.17$ dex. These
corrections are again small compared to the range of $\kappa$ parameters for
stellar systems.

The zero point corrections to the $\kappa$ parameters are summarized in
Equations~5~to~10.  For spiral and irregular galaxies, the $\kappa$-parameters
are:

\begin{equation}
\kappa_1 \quad = \quad (\log {V_{rot}}^2 \, + \, \log r_e)/\sqrt{2}
- 0.21 \, ,
\end{equation}

\begin{equation}
\kappa_2 \quad = \quad (\log {V_{rot}}^2 \, + \,2 \log I_e \, - \,
                                               \log r_e)/\sqrt{6} -0.12 \, ,
\end{equation}

\centerline{and}

\begin{equation}
\kappa_3 \quad = \quad (\log {V_{rot}}^2 \, - \, \log I_e \, - \,
                                               \log r_e)/\sqrt{3} - 0.17.
\end{equation}

\noindent For all other stellar systems (except DHGs), which require 
conversion of $\sigma_e$ to $\sigma_c$, the equations become:

\begin{equation}
\kappa_1 \quad = \quad (\log {\sigma_e}^2 \, + \, \log r_e)/\sqrt{2}
+ 0.11 \, ,
\end{equation}

\begin{equation}
\kappa_2 \quad = \quad (\log \sigma_e^2 \, + \,2 \log I_e \, - \,
                                               \log r_e)/\sqrt{6} + 0.06 \, ,
\end{equation}

\centerline{and}

\begin{equation}
\kappa_3 \quad = \quad (\log \sigma_e^2 \, - \, \log I_e \, - \,
                                               \log r_e)/\sqrt{3} + 0.09.
\end{equation}

\subsection{The Listing of the Data}
\label{kap1-4}

Table~1 lists the data for our sample (the first page only is printed for the
journal).  The full table is electronically available in a convenient ASCII
format through this journal, the Astrophysical Data Center (ADC), and via 
anonymous ftp from samuri.la.asu.edu (129.219.144.156; cd to pub/kappaspace)

Column 1 gives the common name for the object (N = NGC; U = UGC; E = ESO
number; M = MCG number; ZW = Zwicky catalog number; ABELL for Abell Cluster;
GP for Nolthenius group number). Column 2 gives an internal identifying
number. In Column 3 we give a numerical code for the type of object (the
following number in parentheses are the number of objects of each type in the
catalog): 1 for anisotropic gE or cE galaxies (41); 2 for isotropic gE, cE
galaxies or bulges of S0s (51); 3 for isotropic dE galaxies (3); 4 for
anisotropic dE galaxies (8); 5 for 7~Samurai elliptical galaxies with no
isotropy measure (292); 7 for dSph galaxies with no isotropy measure (5); 10
for Sa galaxies (62); 11 for Sab--Sb galaxies (106); 12 for Sbc galaxies (89);
13 for Sc galaxies (129); 14 for Scd--Sdm galaxies (47); 15 for Sm galaxies
(33); 16 for Irr galaxies (45); 18 for the galaxies Malin 1 and Malin 2 (2);
20 for spiral-rich galaxy groups with only 3 galaxies (56); 21 for spiral-rich
galaxy groups with 4 galaxies (17); 22 for spiral-rich galaxy groups with
between 5 and 8 members (32); 23 for spiral-rich galaxy groups with between 9
and 18 members (9); 24 for spiral-rich galaxy groups of 19 members or more
(4); 30 for E-rich groups with 3 galaxies only (12); 31 for E-rich groups with
4 galaxies (20); 32 for E-rich groups with between 5 and 8 galaxies (14); 33
for E-rich groups with between 9 and 18 galaxies (3); 34 for E-rich groups
with between 19 and 113 galaxies (3); 35 for the 16 rich clusters sampled by
Schaeffer et al. (1993) (16); and 40 for the globular clusters (51).

Column 4 gives the distance in units of Mpc ($H_0 = 50$ km sec$^{-1}$
Mpc$^{-1}$).  Column 5 gives the log of the original characteristic internal
velocity ($\sigma_c$ for gE, cE, Bulges and 7~Samurai ellipticals; $V_{rot}$
for spirals and irregulars ($\rm \log V_{\rm char}$, obs); $\sigma_e$ for 
everything else).  Column 6 gives either the observed value of 
$\log \sigma_c$ or the transformed value as discussed in \S~\ref{kap1-2} 
($\rm \log sigma_c$, used).  Columns 7 and 8 give, respectively, the values
of $\log r_e$ in kpc and $\log I_e$ in L$_\odot$ pc$^{-2}$ used to calculate
the $\kappa$ parameters.  These B-band photometric parameters are corrected
for Galactic extinction and internal extinction, usually by the original
source of data (\S~\ref{kap1-1}).  No K-corrections or cosmological
corrections have been applied.

Column 9 gives either the reddening-corrected and K-corrected $B-V$ color or,
in the case of galaxy groups and galaxy clusters, the fraction of E$+$S0 giant
galaxies.  Columns 10, 11 and 12 give the $\kappa$ parameters. Columns 13 and
14 give the quantities $\delta_{2:1}$ and $\delta_{3:1}$ defined in
\S~\ref{kapgal-7}.  Columns 15, 16 and 17 give effective mass, $\log M_e$,
effective B-band luminosity, $\log L_e$ and effective B-band mass-to-light
ratio, $\log (M_e/L_e)$ in solar units.  Column 18 gives the equivalent virial
temperature of the object within the effective radius, $\log T_e$. Lastly,
Column 19 gives the estimated baryon number density log $n_{bary,e}$
(cm$^{-3}$).  The equations to generate many of the more familiar quantities
derived from the $\kappa$ parameters are given in Appendix~\ref{app-1}.

\section{Galaxies in $\kappa$-Space}
\label{kapgal}

\subsection{Spiral and Elliptical Galaxies in the Virgo and Coma Clusters}
\label{kapgal-1}

Figure~1 shows the distribution in $\kappa$-space for 48 gE in the Virgo and
Coma clusters (cf. Figure~1 of $\rm B^2F1$) plus 15 spiral galaxies, Hubble
types Sa through Sc/d, in Virgo.  (These graphs differ slightly from those
shown in $\rm B^2FN$ owing to the $K_1$ velocity corrections made to the
spiral galaxy $\kappa$ parameters here.) We have previously shown that, when
distances are defined as they are here, the Virgo and Coma gE samples define a
very thin fundamental plane. Figure~1 is a two-dimensional fold-out of
three-dimensional $\kappa$-space. By viewing the points projected onto all
three planes, one can reconstruct their distribution in three dimensions.  Two
particular features of these diagrams are highlighted.

The FP defined by gEs in Virgo and Coma is given by the dark solid line in
$\kappa_1$--$\kappa_3$ (where the FP is viewed edge-on by design).  Its
equation is:

\begin{equation}
\kappa_3 = 0.15 \kappa_1 + 0.36 \, ,
\end{equation}

\noindent
or, in solar units 

\begin{equation}
\log (M_e/L_e) = 0.184 \log M_e - 1.25 \, .
\end{equation}


\noindent Equation~12 is the conventional definition for the DHG FP in terms
of mass-to-light ratio: $M_e/L_e \propto M_e^{0.184}$.  The FP projects onto
$\kappa_2$--$\kappa_3$ as a thick band, the midsection of which is
represented by the light solid line in this projection, the dotted lines
indicating the upper and lower boundaries of the DHGs. The FP projects nearly
face-on in $\kappa_1$--$\kappa_2$, where the distributions of DHGs and
spirals are both very broad.  The short dark lines drawn in each panel of
Figure~1 and subsequent figures represent the effect of distance errors of
$\pm$30\%. Distance errors owing to real peculiar motions of galaxies up to
this amplitude are the dominant source of observational scatter in all
diagrams save Figure~1.

As is evident, the physical properties of the spiral galaxies in Virgo occupy
a similar distribution in $\kappa$--space to the elliptical galaxies.  There
is much overlap in $\kappa_1$--$\kappa_2$, with offset but parallel
distributions in the other two projections.  The Virgo spirals thus define a
second fundamental plane that is nearly parallel to the DHG FP but shifted to
lower mass-to-light ratios by about a factor of two at fixed mass.

\subsection{The Zone of Exclusion (ZOE) in $\kappa_1$--$\kappa_2$}
\label{kapgal-2}

The diagonal dotted line in the $\kappa_1$--$\kappa_2$ plane in Figure~1
delineates the empirically-determined ``zone of exclusion'' (hereafter called
ZOE) for DHG galaxies ($\rm B^2F1$).  Its equation is

\begin{equation}
\kappa_1 + \kappa_2 \le 8 \, 
\end{equation}

\noindent 
The zero point here is adjusted slightly from the value 7.8 used in $\rm B^2F1$
to make the exclusion line exclude essentially all stellar systems including
rich clusters (see below). Note that the Virgo spirals obey the same ZOE as
the DHGs. This is a theme we will return to later:  the ZOE is {\it universal}
for the present sample of galaxies and clusters.

We define an effective volume density $j_e \equiv L_e \times (4/3 \pi
r_e^3)^{-1}$, which in the current system of units is $0.75 \times 10^{-3}
I_e/r_e$ L$_\odot$ pc$^{-3}$.  Equation~13 can then be recast as

\begin{equation}
\log M_e + 0.73~ \log j_e \le 10.56 \, ,  
\end{equation}

\noindent or approximately

\begin{equation}
j_e \le {\rm const.} \times M_e^{-4/3} \, .
\end{equation}

\noindent In other words, the maximum allowed luminosity volume density for
collapsed stellar systems scales as $M_e^{-4/3}$.  The significance of this
condition will be made clear when we examine the distribution of all stellar
systems relative to the ZOE.

\subsection{Dynamically Hot Galaxies}
\label{kapgal-3}

Figure~2 plots the distribution of all DHGs in $\kappa$-space. This
distribution has been extensively discussed elsewhere ($\rm B^2F1$; Bender,
Burstein, \& Faber 1993a,b, 1995).  In the present figures we divide the DHGs
into six subsets for plotting purposes: isotropic and anisotropic giant Es,
compact Es and bulges; isotropic and anisotropic dwarf ellipticals; giant Es
from Faber et al. (1989) and dwarf spheroidals. These categories are defined
and explained in B$^2$F1.

DHGs divide into two major families that are most clearly seen in
$\kappa_1$--$\kappa_2$:  giant galaxies (upper right), which we call the
Gas-Stellar Continuum in B$^2$F1, and dwarf galaxies (lower left).  This is
analogous to the division of DHG systems into giants and dwarfs found by
Kormendy (1988). Within the Gas-Stellar Continuum, more massive giant galaxies
tend to have lower surface brightness (lower $\kappa_2$), while more massive
dwarf galaxies tend to have higher surface brightness.  Thus, in
$\kappa_1$--$\kappa_2$ the two families are at approximately right angles to
one another (Figure~2).

In B$^2$F1, we proposed the name Gas-Stellar Continuum because several
properties vary along the giant sequence in a way that suggests that a varying
fraction of gas (vs. stars) was involved in the last major merger.  For
example, giant ellipticals rotate slowly, are typically anisotropic in their
velocity distributions, and tend to have boxy isophotes.  Less massive
ellipticals tend to be isotropic in velocity dispersion, rotate rapidly and
have disky isophotes (Davies et al. 1983; Bender 1988; Kormendy \& Bender
1996). Bulges continue the trend from less massive ellipticals.  As discussed
in $\rm B^2F1$ (see also references cited therein), these data suggest a
merger formation picture in which the latest mergers that formed the smaller
galaxies contained considerable gas, whereas those that formed the more
massive ones had progressively more stars. Recently measured central density
profiles (Faber et al. 1997) are consistent with this view.  For brevity, in
what follows we will refer to the giant DHG galaxies that comprise the
Gas-Stellar Continuum as ``giant ellipticals," or gEs, despite the fact that
this sequence also contains compact Es and bulges.

The dwarf family of DHGs is offset from the giant family to higher $M/L$ in
$\kappa_1$--$\kappa_3$, and this trend becomes extreme among the dSph, which
are heavily dominated by DM.  It is likely that they have lost baryons by
galactic winds (Dekel \& Silk 1986), ram-pressure stripping (Lin \& Faber
1983), or some other means.  $\rm B^2F1$ discussed ways in which the dwarf E 
family might evolve to or from the giants, but options are limited.
Present-day dwarf Es cannot be merged to produce present-day gEs as their
stellar populations are too different ($\rm B^2F2$). Conversely, if dE and
dSph are formed from other Hubble objects via either galactic winds or
ram-pressure stripping, B$^2$F1 showed that their progenitors are not visible
among the other DHGs. However, they do seem to be compatible with baryon mass
loss from late-type spiral and/or Irr progenitors, as discussed below.

\subsection{Spiral Galaxies of Types Sa, Sab and Sb}
\label{kapgal-4}

Figure~3 shows the distribution within $\kappa$-space for spiral galaxies of
Hubble types Sa to Sb (RC3 type numbers 1 to 3). This figure illustrates
several points:

1)  Early-type spiral galaxies are distributed similarly to gEs but offset
slightly in each $\kappa$-space projection.  Sa--Sb galaxies lie farther from
the ZOE than gEs.

2)  Sa--Sb galaxies define their own FP relation in $\kappa_1$--$\kappa_3$
that is nearly parallel to that of gEs but slightly offset to lower values of
$M/L$ by about 0.3 dex. This is consistent with the Virgo cluster spirals
(Figure 1).

3) There is a slight tilt to this plane in $\kappa_2$--$\kappa_3$ compared
to the nearly level distribution of DHGs in Figure~2.  While barely
visible here, this tilt becomes progressively more pronounced for later
Hubble types.  We will find in \S~\ref{kapgal-8} that it is the key 
that yields the TF relation.

\subsection{Spiral Galaxies of Types Sbc and Sc}
\label{kapgal-5}

Figure~4 shows the distribution within $\kappa$-space for spiral galaxies of
Hubble types Sbc and Sc (RC3 type numbers 4 and 5).  The march of late-type
galaxies away from the ZOE continues with this sample.  The locus of Sbc--Sc
galaxies within $\kappa_1$--$\kappa_2$ parallels that of the earlier Hubble
types, including the gEs and bulges. Within $\kappa_1$--$\kappa_3$ and
$\kappa_2$--$\kappa_3$, the Sbc--Sc's generally follow the Sa--Sb's.

When these data are merged with the Sa--Sb's, we can see that early
disk-dominated Hubble types Sa--Sc form a fundamental plane that is offset
from and tilted slightly with respect to the DHG FP.  The principal difference
as a function of Hubble type is not the plane that each type defines, but
rather the {\it distribution of galaxies within the plane}: later Hubble types
are shifted to slightly lower masses and to lower surface brightnesses at a
given mass.

\subsection{Spiral Galaxies of Types Scd through Irr}
\label{kapgal-6}

Figure~5 shows the distribution within $\kappa$-space for spiral galaxies of
types Scd to Irr (RC3 type numbers 6 through 10).  Included in this diagram
are the very large, very low surface brightness galaxies Malin 1 and Malin 2.
In contrast to the earlier-type spiral galaxies and the gEs, these late type
galaxies show essentially no correlation in either $\kappa_1$--$\kappa_2$ or
$\kappa_1$--$\kappa_3$. However, their tilt in $\kappa_2$--$\kappa_3$ is now
very pronounced in the sense of having lower $\kappa_2$ values for higher
$\kappa_3$.  This tilt has erased all hint of correlation in the usual
edge--on FP projection ($\kappa_1$--$\kappa_3$).  In essence,
$\kappa_2$--$\kappa_3$ replaces $\kappa_1$--$\kappa_3$ as the proper FP
edge--on projection for late-type galaxies.  Even though these galaxies show
no correlation in $\kappa_1$--$\kappa_3$, their median $M/L$ is still
coincident with the DHG fundamental plane.

The properties of the two very large spiral galaxies Malin 1 and Malin 2 place
them in unusual locations. Malin 2 lies close to the most massive end of the
DHG FP in $\kappa_1$--$\kappa_3$, lies among late-type spirals in
$\kappa_2$--$\kappa_3$, but has much higher mass than typical spirals. Malin 1
is even more massive with still lower surface brightness, larger radius and
larger $M/L$. The positions of Malin 1 and Malin 2 within $\kappa$-space
become more understandable when we discuss the properties of galaxy groups in
\S~\ref{kapgp}.

\subsection{Quantifying the Interrelationships Among Galaxies in 
$\kappa$-Space}
\label{kapgal-7}

\subsubsection{The Distribution of Galaxies Relative to the ZOE}
\label{kapgal-7-1}

The quantity 

\begin{equation}
\delta_{2:1} \equiv \kappa_1 + \kappa_2 - 8
\end{equation}

\noindent measures the distance of a galaxy from the ZOE in
$\kappa_1$--$\kappa_2$. Figure~6 shows histograms of $\delta_{2:1}$ divided by
galaxy type. Median values, $\delta_{2:1}\,(med)$, for each type of stellar
system are shown in the figure. These histograms reinforce what we saw in the
individual $\kappa$ diagrams. The median values of $\delta_{2:1}$ show a
steady progression along the Hubble sequence away from the ZOE, starting with
$\delta_{2:1} \, (med) = -0.70$ for gEs, --0.90 for Sa galaxies, and through
to -3.16 for Irr galaxies.  For earlier Hubble types gE to Sbc, the histograms
are reasonably Gaussian in shape, with similar widths and well defined means
and medians. Sc galaxies have a bit wider histogram distribution, but
$\delta_{2:1}$ is still reasonably Gaussian.  From Scd to Irr galaxies, the
distributions become progressively wider, with possibly large wings.

What if we had {\it not} made the $K_1$ correction to convert $V_{rot}$ to
$\sigma_c$, which amounts to $\Delta \delta_{2:1} = -0.36$?  In that case, Sa,
Sab and Sb galaxies would be closer to the ZOE than the gEs, and the steady
downward march through {\it all} Hubble types would be broken.  The
velocity-related $K_1$ correction, though small, seems necessary to properly
rank-order the Hubble types.

\subsubsection{The Distribution of Galaxies Relative to the Fundamental Plane}
\label{kapgal-7-2}

An analogous quantity 

\begin{equation}
\delta_{3:1} \equiv \kappa_3 - 0.15 \kappa_1 - 0.36
\end{equation}

\noindent measures the vertical distance of a galaxy in $M/L$ from the
DHG-defined fundamental plane (Equation~11). Figure~7 plots histograms of
$\delta_{3:1}$ for Hubble types divided as in Figure~6.  Unlike Figure~6,
Hubble types here march first in one direction and then back again. Early
Hubble types, gE--Sbc, move to lower $M/L$ with advancing type.  This probably
reflects a greater number of young stars with later Hubble type.  However,
this trend begins to reverse at Sc, with later types marching back to {\it
higher} $M/L$.

Several authors (Tinsley 1981, Faber 1982a, Verheyen 1997) have remarked on the
higher $M/L$ values of very late type galaxies in the B passband. However,
this is the first time to our knowledge that the trend has been detected to
set in as early as Sc.  There are at least two factors that could contribute
to this trend: 1) a tendency for more baryons to be in non-stellar form (i.e.,
H~I and H$_2$ gas) in later types, and 2) more DM relative to baryons within
the optical radius.  We return to this issue briefly in \S\ref{kaphcm-1}.

\subsubsection{The Distribution of dE's versus Scd--Irr Galaxies}
\label{kapgal-7-3}

We have seen that each Hubble type occupies a distinct region of the
$\kappa_1$--$\kappa_2$ plane, with some overlap between neighboring Hubble
types.  Exceptions are the dEs and Scd--Irr galaxies, whose loci lie squarely
on top of one another.  This means that their masses and radii are similar,
though the mass-to-light ratios of dEs are higher by roughly a factor of five
at a given mass.  In $\rm B^2F1$ we noted that dEs do not seem to be
evolutionarily linked to giant DHGs because known evolutionary processes ---
tidal stripping, ram-pressure stripping, mergers, galactic winds --- all move
them in wrong directions in $\kappa_1$--$\kappa_2$.  However a structural link
with Scd-Irr's seems plausible.  If gas were removed from small late-type
galaxies, causing star formation to slow or cease, and if the remaining stars
faded, the general location and high $M/L$'s of the dEs in $\kappa$-space
might be explained.  Several authors have suggested mechanisms that might
strip gas from the shallow potential wells of small late-type galaxies via
stripping, winds or ``harassment'' in clusters of galaxies (e.g., Lin \& Faber
1983; Dekel \& Silk 1986; Moore et al. 1996).

\subsubsection{A Dissipation Strip in $\kappa$-Space?}
\label{kapgal-7-4}

As noted in $\rm B^2F1$, isotropic DHG galaxies (mostly lower-luminosity gEs)
tend to be found in a narrow strip in $\kappa$-space parallel to the
$\kappa_2$ axis and with boundaries $\kappa_1 = 2.7$ ($M_e \approx 10^{9.5} \,
M_\odot$) and 4.3 ($\approx 10^{11.75} \, M_\odot$). Interestingly, among the
511 spiral galaxies in our sample, only 21 galaxies, all of Hubble type Sdm or
Irr, lie more than 0.2 dex to the low mass side of this range, and only 4
spiral galaxies lie 0.2 dex or more to the high side.  In other words, nearly
all disk galaxies lie in the same mass range as the {\it isotropic} DHGs. 
Isotropy among DHGs implies flattening by high rotation.  This, plus their
high central densities (Faber et al. 1997) and probable oblate shapes
(Tremblay \& Merritt 1996) points to significant gaseous dissipation.  Perhaps
this strip in $\kappa$-space, $M_e = 10^{9.5} - 10^{11.75} \, M_\odot$, is the
range of mass in which baryonic dissipation results in a flattened, rotating
galaxy.

\subsection{The Tully-Fisher Relation in $\kappa$-Space}
\label{kapgal-8}

A generalized TF relationship can be expressed as $\log L_e = A_C \log
V_{rot}$, where $A_C$ represents the color-dependent exponent in the standard
TF power law.  Using the relations defined in Appendix~\ref{app-1} we can 
rewrite this as

\begin{equation}
[\frac{4 - A_C}{\sqrt{2}}] \, \kappa_1 - \frac{A_C}{\sqrt{6}}
\, \kappa_2 - \frac{A_C + 6}{\sqrt{3}} \, \kappa_3 = constant \, .
\end{equation}

\noindent This is the equation of a plane in $\kappa$-space.  We have
already discussed the planar distribution of spirals and likened it to the
fundamental plane for DHGs, which it resembles except for a slight tilt.

If $A_C = 4$ (i.e., $L_e \propto V_{rot}^4$), the coefficient of $\kappa_1 =
0$ and we get the simple relation $\kappa_3 \propto -\frac{1}{2.5 \sqrt{2}}
\kappa_2 = -0.283 \kappa_2$.  This plane is parallel to the $\kappa_1$ axis and
projects with minimal scatter onto $\kappa_2$--$\kappa_3$.  We have already
noted the small scatter of spirals and Irr's in $\kappa_2$--$\kappa_3$, so the
real spiral fundamental plane cannot be far from this line.  If $A_C = 3$,
$\kappa_1 - \sqrt{3} \kappa_2 - 3 \sqrt{6} \kappa_3 = const$, which is a
tilted plane that does not project perfectly to a line on any pair of $\kappa$
axes.

The TF relation defined by our spiral and irregular galaxy sample is given in
Figure~8.  Part of the large scatter in this diagram comes from ignoring
peculiar motions (cf. Faber \& Burstein 1988).  Nevertheless, it is evident
that a relation of the form $-2.5 \log L_e \propto -7.5 \log V_{rot}$ defines
this TF relation well, which implies $A_C = 3$. A formal fit yields $-2.5
\log L_e \propto -7.6 (\pm 0.23) \log V_{rot}$ for all spirals and irregulars
taken together, with coefficients of 7.48 $(\pm 0.1)$ for Hubble types Sa-Sc,
and 8.70 $(\pm 0.35)$ for later Hubble types.

In other words, the fundamental plane for DHGs and the TF relationship for
spiral galaxies are basically the same thing --- each is the virial plane for
its galaxy type, illustrating a family of self-gravitating objects in
dynamical equilibrium and illuminated by stellar populations with well behaved
mass-to-light ratios that vary as power-law functions of the basic structural
variables (cf. Faber et al. 1987).  This fact has been implicit since Faber \&
Jackson derived $L \propto \sigma^4$ for gE galaxies and Aaronson et al. (e.g.,
Aaronson \& Mould 1986) derived $L \propto V_{rot}^4$ for spiral galaxies.
What was not clear until galaxies were graphed together in $\kappa$-space is
that the physical properties of spirals are continuous with those of DHGs. 
Indeed, DHGs define but one fundamental plane within $\kappa$-space;
early-type spirals define another, and late-type spirals yet a third. In
reality, there is probably a continuum of fundamental planes, just as we know
there is a continuum of Hubble types, but we simplify them here to just three.

{\it If} observational errors and inconsistencies in the defined system of
structural parameters are not important, and if galaxies are truly in
gravitational equilibrium, then the existence of multiple FPs {\it must}
reflect differences in mass-to-light ratios.  This becomes obvious when
one considers the {\it purely dynamical} version of fundamental plane space
($\sigma$,$r$,$\Sigma$) where mass surface density ($\Sigma$) replaces surface
brightness.  Since $\Sigma \propto M/r^2$ and since the virial theorem says
that $M = V^2 r/G$, it is clear that there can be only one such plane. 
However, we cannot observe $\Sigma$ independently of $\sigma$ and $r_e$, so we
replace it with the next best parameter, $I_e$, which we {\it can} observe. 
The two are related by $I_e \propto \Sigma (M/L)^{-1}$, which introduces a new
wild card, mass-to-light ratio. It is the systematic variation in $M/L$ as a
function of $\sigma$ and $r$ for the different Hubble types that creates the
multiple fundamental planes.

If we derive a three parameter, ``FP--like'' fit for the spiral and irregular 
galaxies we get $\log V_{rot} \propto 0.61 (\pm 0.02) r_e + 0.39 (\pm 0.02)
I_e$ for the whole sample, with coefficients (0.85$\pm$0.05, 0.57$\pm$0.03)
for spirals Sa-Sc, and (0.54$\pm$0.03, 0.34$\pm$0.02 for late-type galaxies.  
Errors were estimated using a bootstrap method using different subsamples of 
half the data at a time.  These coefficients are to be compared with the
usual values for the gE FP, (0.72$\pm$0.07, 0.67$\pm$0.07), derived, as
for example, by Faber et al. (1987).  As we can see, the FP defined by
the early-type spiral galaxies is reasonably consistent, within errors, with
that of gE galaxies.  In contrast, the FP defined by late-type galaxies
is somewhat flatter.  In both cases, the fits confirm what we see in the
figures.

\section{Galaxy Groups, Galaxy Clusters and Globular Clusters in 
$\kappa$-Space}
\label{kapgp}

\subsection{Groups and Clusters}
\label{kapgp-1}

To examine the $\kappa$ parameters of galaxy groups and clusters, we shift the
center of the volume to larger masses (higher $\kappa_1$), lower mean surface
brightnesses (lower $\kappa_2$) and higher mass-to-light ratios (higher
$\kappa_3$). Figure~9 shows such a plot of $\kappa$ parameters for galaxy
groups and clusters together with the giant spirals Malin~1 and Malin~2.

Galaxy groups are harder to identify uniquely than galaxies or galaxy
clusters.  Care must be taken to verify that groups defined in redshift space
have properties that are similar to the ``pure'' groups one would have
identified without contamination, redshift smearing, and projection biases.
This issue is addressed using N-body simulations in Appendix~\ref{app-2}.

As pointed out by Schaeffer et al., rich clusters define a fundamental plane
that is parallel to but offset from that of DHGs. This can be seen in
Figure~9, where their projection in $\kappa_1$--$\kappa_3$ is remarkably tight
(large black dots).  Galaxy groups define a second plane that is canted at
a relatively steep angle to that of both the rich clusters and the giant 
ellipticals, bridging the two parallel planes between the high-mass galaxy 
end and the low-mass cluster end.

To quantify this picture, we calculate values of the $M/L$ residual
$\delta_{3:1}$ (Equation~17) for rich clusters and galaxy groups, the latter
divided between spiral-rich and elliptical-rich groups  and between groups
with 10 or more galaxies versus 9 or fewer.  Histograms of $\delta_{3:1}$ are
shown in Figure~10, with the histogram for gE galaxies from Figure~7 included
for reference. Figures~9 and 10 suggest some important conclusions about the
physical properties of different kinds of groups and their relationship to
galaxies:

1)  The fundamental plane for rich clusters is parallel to that of DHGs but
offset in $\kappa_3$ by 0.54 dex, or nearly a factor of 10 in mass-to-light
ratio.  Roughly 1/3 of this effect is due to the higher fraction of X-ray gas
in clusters (Mushotzky 1991), which does not shine in B light, and the rest is
due to the greater DM enclosed in clusters. Remarkably, the 16 rich clusters
define a fundamental plane that is, if anything, {\it narrower} than the plane
defined by bright elliptical galaxies.

2) The fundamental plane for galaxy groups is not a plane but a curved
surface.  For populous E-rich groups, this surface is nearly coincident with
the plane for rich clusters, tilting down to lower $M/L$ values for
sparsely-populated S-rich groups. Yet rich groups, poor groups, E-rich groups
and S-rich groups all define $\kappa$-space distributions that are seen
edge-on in $\kappa_1$--$\kappa_3$.  The 10 groups that lie significantly high
in $M/L$ fall into two categories:  groups within the Local Supercluster whose
distances are underestimated at a level $>$100\% by large-scale motions; and
probable spurious and/or under-sampled distant groups with very large velocity
dispersions. It is possible that the abnormally high $M/L$ for these few 
groups are artifacts.

3)  The low-mass end of the distribution of galaxy groups in $\kappa$-space
intersects the high-mass end of the DHG FP for individual galaxies. It seems
strange that the $M/L$ for groups, which enclose DM, should be the same as
that for galaxies, which are dominated by stars. However, the masses of
low-mass groups are potentially afflicted by bigger observational errors due
to projection effects and small-number statistics. Note that errors in the 
virial masses tend to move groups along a direction roughly parallel to the 
FP for low mass groups, as shown by the arrow in $\kappa_1$--$\kappa_3$ in
Figure~9.  The effect of errors on the
parameters of small groups is discussed further in Appendix~\ref{app-2} using 
N-body simulations.  Errors may contribute some portion of the different tilt 
of low-mass groups.  However, even factor of two errors in $\sigma_e$ are
not nearly enough to produce the full distribution of groups, so we conclude
that most of the effect appears to be real.

4)  Groups and clusters obey the same $\kappa_1 = -\kappa_2 +8$ ZOE defined
by the DHGs. The richest clusters appear to have high enough mean surface
brightnesses to bring their $\kappa_2$ values up to this line, while the N93
groups show an upper boundary roughly parallel to, but offset below this
line.  This difference may be a richness effect caused by sampling different
volumes --- the Schaeffer et al. sample is drawn from a much larger volume
than the CfA1 groups, and it is likely that the CfA1 volume is too low to
include any large clusters.

5)  Within $\kappa_1$--$\kappa_2$, rich clusters and rich groups define a
loose relationship that projects nearly perpendicularly to the ZOE,
paralleling the distribution of dwarf DHGs and Scd--Irr's but at greater
masses and fainter surface brightnesses.  Like the late-type spiral galaxies,
the distribution of small galaxy groups in $\kappa_1$--$\kappa_2$ is
amorphous. Some of the scatter for the small groups may stem from small-number
statistics and projection effects.

6) Most E-rich groups (circles) are separated from S-rich groups (triangles)
by a line that is roughly parallel to, but offset from, the ZOE in
$\kappa_1$--$\kappa_2$.  Just as galaxies march away from the ZOE at later
Hubble types (Sec~\ref{kapgal-7-1}), so groups march away as the fraction of 
their late-type galaxies increases.

7) The ``dissipation strip'' previously defined for galaxies
(\S~\ref{kapgal-7-4}) ends at $\kappa_1 \sim 4.3$.  This is also a reasonable
lower bound to the mass of groups, as expected since groups here are 
explicitly identified by only their galaxies.

8)  The $\kappa$ parameters of the galaxy Malin 1 place it at the low-mass end
of groups.  Malin 2 is similar but closer to the gap between galaxies and
groups.  The structural parameters of these galaxies suggests that they may be
``failed groups,'' as we termed them in $\rm B^2FN$. By that, we mean systems
in which baryonic dissipation and infall was so slow that the process of
galaxy formation was interrupted by the formation of a ``group." Such events
are evidently rare.

\subsection{Globular Clusters and an Overall View of $\kappa$-Space}
\label{kapgp-2}

Globular clusters are included in the catalog because they are the oldest
objects in our Galaxy, with some perhaps having formed ``primordially" before
the Galaxy as a whole collapsed.  Globulars are also self-gravitating, and it
is our desire to consider all self-gravitating stellar systems together.  To
view all stellar systems in one graph, we must collapse the scale down to such
an extent that, for any reasonable symbol size, the points merge together to
form an inky blob.  Such a view of $\kappa$-space is given in Figure~11, which
shows globular clusters relative to those of other stellar systems. They are
separated from larger systems by a large gap, being denser and much less
massive. Previous work (e.g., Schaeffer et al. 1993; Djorgovski 1995) has
shown that the physical properties of globular clusters tend to project them
along a fundamental plane parallel to but offset from that of DHGs.  This is
shown in Figure~10, which shows the histogram of $\delta_{3:1}$ values for the
51 globular clusters in our sample.

Globular cluster distances are determined in a manner that is largely
independent of the assumed Hubble constant.  If the assumed distance scale for
globular clusters is systematically different from our cosmic scale (based on
$H_0 = 50$ km sec$^{-1}$ Mpc$^{-1}$), the difference will show up as a spurious
offset most evident in $\kappa_1$--$\kappa_3$. The mean value of
$\delta_{3:1}$ for the globulars is +0.26 dex Figure~10).   If $H_0 = 80$ km
sec$^{-1}$ Mpc$^{-1}$, $\delta_{3:1}$ would be reduced to +0.14 dex (cf. Nieto
et al. 1990).

The distribution of globular clusters in $\kappa$-space seems more linear than
planar.  This line is most clearly seen in $\kappa_1$--$\kappa_2$ but can also
be seen in $\kappa_2$--$\kappa_3$.  The $\kappa$-space distribution of
globular clusters is more line-like than that of any other stellar system.

Many formation mechanisms have been suggested for globular clusters, including
collapse from primordial fluctuations (Peebles \& Dicke 1968; Rosenblatt et
al. 1988), cooling flow thermal instabilities (Fall \& Rees 1985),
merger-induced shocks (Holtzman et al. 1992; Ashman \& Zepf 1992), and
molecular cloud collapse within galaxies (Searle \& Zinn 1978). Evidence from
detailed examination of Galactic globular clusters suggests that, regardless
of formation mechanism, only those clusters with long evaporation times and
large perigalacticon distances survive today.  Evaporation and potential
shocking whittle away at an initial distribution, selecting as survivors only
those clusters within a narrow range of {\it radii} (Gnedin \& Ostriker 1997).
This is consistent with the distribution of globular clusters in Figure~11,
which is elongated and tilted with a zero point that corresponds to the
predicted radius of 1-10 pc (cf. Figure~15).

\section{$\kappa$-Space and Hierarchical Clustering}
\label{kaphcm}

We turn now to a comparison of the topology of stellar systems in
$\kappa$-space with the predictions of structure formation by hierarchical
clustering.  Figure~11 can be summarized by saying that the $\kappa$-space
distributions of common stellar systems are viewed mostly edge-on in
$\kappa_1$--$\kappa_3$, mostly face-on in $\kappa_1$--$\kappa_2$, and at
intermediate angles in $\kappa_2$--$\kappa_3$. Each kind of stellar system
defines a slightly different FP in this diagram, though they are all related. 
We suggest the term {\it cosmic metaplane} for this ensemble of continuous
interlocking, and mainly parallel fundamental planes.

In retrospect, the existence of the metaplane is not surprising. We know from
DHGs that any constraint of the form $M/L = f(M)$ will force a plane that is
edge-on in $\kappa_1$--$\kappa_3$.  The existence of the metaplane reflects
the empirical fact that $M/L$ is closely constrained by an object's mass. In
the case of galaxies, this implies the known fact that galaxies of a similar
Hubble type and size make stars in similar fashion. In the case of galaxy
clusters, we can now add an additional requirement that clusters of a given 
mass have similar stellar baryon fractions.  The physics behind these 
constraints is still unknown, especially the physics of star formation. 
However, the net result is that $M/L$ varies by at most 100 over a mass range 
of 10$^9$ M$_\odot$, and the range in $M/L$ at any given mass is less than 10.
It is this weak overall variation and local tightness in $M/L$ that creates the
metaplane.

However, for the purpose of understanding structure formation, the face-on
view of the metaplane as seen in $\kappa_1$--$\kappa_2$ is more useful than the
edge-on view in $\kappa_1$--$\kappa_3$ because of the information it contains
on radii and densities.  It is clear from Figure~11 that the properties of
galaxies and clusters are highly organized and interrelated in
$\kappa_1$--$\kappa_2$, and there appears to be a wealth of information on how
structure formed. In this section we consider the distribution of objects in
$\kappa_1$--$\kappa_2$ in light of the prevailing paradigm for structure
formation involving hierarchical clustering and merging (hereafter HCM).

\subsection{Galaxies versus Clusters of Galaxies in $\kappa$-Space}
\label{kaphcm-1}

The present comparison to data is similar to the approach taken by BFPR,
expanded to the full 3-D $\kappa$-space and incorporating our newer and more
extensive database for galaxies and clusters.  The top hat spherical
dissipationless collapse model for fluctuations is assumed (Gott \& Rees
1975). If the fractional overdensity of a spherical fluctuation in a sphere of
mass $M$ is $\delta \equiv {\delta \rho}/\rho$ at redshift $z$, then the
equilibrium radius of the collapsed dark halo formed from this fluctuation is
(Faber 1983):

\begin{equation}
R_{dh} = 0.79~ M_{12} \delta^{-1} h^{-2/3} \Omega^{-1/3} (1+z)^{-1}~ {\rm Mpc},
\end{equation}

\noindent where $M_{12}$ is the mass in units of $10^{12}$ M$_{\odot}$. This
expression contains two independent parameters, $\delta$ and $M_{12}$, which,
when varied, generate a two-dimensional family of collapsed objects that
populate the {\it cosmic virial plane}. As previously noted, this is the
parent plane for all the observed individual fundamental planes.  The
injection of different $M/L$ rules for different families of objects at
different locations transforms the virial plane into the collection of
observed, interlocking fundamental planes that we have called the cosmic
metaplane.

The cosmic virial plane is infinite in extent, defined by the ensemble of all
possible collapsed objects of all masses and radii. In the real universe, this
plane is only partially filled by the actual clustering process.  The
challenge of any theory of structure formation is to reproduce the observed
distribution of objects versus mass and radius in this plane, i.e., to
populate the plane with the right numbers of objects in the right places. 
Furthermore, to compare with observations, one must calculate {\it observed}
radii for galaxies, which involves a theory of angular momentum generation and
dissipation for these objects.  Finally, to reproduce the full distribution in
a $\kappa$-space defined by the B passband, a theory for $M/L$ is needed, and
thus a theory for converting baryons into B-band starlight.

We do not possess knowledge of all of these ingredients at this time.
Nevertheless, schematically it appears that the distribution of objects in
$\kappa$-space is broadly consistent with the predictions of HCM.  To see
this, we focus on $\kappa_1$--$\kappa_2$ to take advantage of the radii
information there.  We also omit globular clusters, since how or whether they
formed from primordial density fluctuations is unclear.

Since objects with higher overdensity collapse first in an expanding universe,
clustering starts with the mass that has highest $\delta$.  Realistic density
fluctuation spectra (e.g., Cold Dark Matter and its variants) decline
monotonically with mass, so that clustering starts with small, high-density
seeds of mass $\sim 10^7$ M$_{\odot}$ (BFPR) and progresses to form larger
masses of lower density. At any epoch, one expects to find a broad range of
densities for collapsed objects at fixed mass, reflecting the fact that the
initial overdensity $\delta$ of an individual sphere of mass $M$ can vary
widely.

The observed distribution of objects in $\kappa_1$--$\kappa_2$ is consistent
with this picture.  The broad swath of points from upper left to lower right
represents the collection of objects that have collapsed to date. Clustering
started at the upper left at small masses and is progressing to the lower
right.  The total vertical width of the swath in $\kappa_2$ at fixed mass
$\kappa_1$ corresponds to a factor of about 20 in radius. This is roughly
consistent with expectations for an initial spectrum of Gaussian random
fluctuations, as we show below. The division between galaxies and clusters of
galaxies is a vertical line at $5.6 \times 10^{11}$ M$_{\odot}$ (corresponding
to $\kappa_1 = 4.3$; \S~\ref{kapgal-7-4}).  Since $5.6 \times 10^{11}$ 
M$_{\odot}$ is the mass of a big galaxy and it takes at least three galaxies 
to make a group, the location of this line is not surprising.  

However, in a deeper sense, the division between galaxies and clusters marks
the boundary of baryonic dissipation: galaxies (for the most part) are objects
in which global baryonic dissipation has occurred and to a large extent,
groups and clusters are objects in which it has not. Thus, explaining why this
line is located exactly where it is and why it appears to be {\it vertical},
i.e., at fixed mass, requires a full theory of dissipation, which is beyond
the scope of this paper.  In a general way, the location of this line probably
reflects the well known boundary for gaseous ionized spheres that can cool
before being incorporated into the next level of the merging hierarchy (Rees
\& Ostriker 1977, BFPR).

Galaxies and clusters are separated in $\kappa_1$--$\kappa_2$ with a clear gap
between them.  This is due to baryonic dissipation. To develop a schematic
estimate of its effects, we assume that the dissipated baryonic mass per dark
halo is $M_{bary} = f_{bary} M_{dh}$, where $f_{bary}$ is of order 0.1 (BFPR),
and that the baryonic radius after collapse is given by $r_{bary} = f_{bary}
r_{dh}$. Use of the same factor, $f_{bary}$, in both formulae ensures that the
rotational velocity of the collapsed baryons, $V_{rot}$, will approximately
equal $V_{dh}$.  This is correct for collapse within an isothermal dark halo
and also fits our current models of galaxy rotation curves.  We further assume
that, for galaxies, the effective mass, $M_e$, is the same as $M_{bary}$. 
With these assumptions, $M/L$ decreases by a factor of $f_{bary}$, while
surface brightness $I_e$ increases by $f_{bary}^2$.

These estimates enable us to undo the effects of baryonic infall and
reconstruct what galactic dark halos would look like if the presently visible
light were distributed like the DM. Figure~12a shows an enlargement of
$\kappa$-space focusing just the $kappa_1$--$\kappa_2$ plane for galaxies,
groups and clusters.  The arrow labeled ``dissipative infall'' represents the
above infall model. Objects shrink in both mass and radius as a result of
baryonic infall, and thus move in both $\kappa_1$ and $\kappa_2$.

Figure~12b attempts to reconstruct the invisible dark halos of galaxies by
sliding all galaxy points back along the above infall vector, assuming that
$f_{bary} = 0.1$.  If hierarchical clustering is continuous across the
boundary from dissipational to dissipationless collapse, we expect the
reconstructed dark halos of galaxies to merge smoothly with those of groups
and clusters.  The reconstructed galaxy halos have that property, confirming
that the original gap between the two families was the result of dissipation.

Note that dissipative infall broadens the total range of masses that can be
plotted in Figure~12.  Many dark halos of galaxies have lost their separate
identities after spawning galaxies by merging with other halos (the so-called
``overmerging'' phenomenon, Katz \& White 1993). Visible galaxies also merge
but much more slowly since their cross-sections are reduced by baryonic
dissipation.  Thus, Figure~12b, which plots both halo populations together, is
broader in mass than the population of halos at any one time --- many of the
reconstructed galaxy halos no longer exist because they have since merged with
other halos to form the dark halos of groups and clusters.

Is the general slope of the ZOE consistent with HCM?  Figure~12b shows that a
major factor in the slope is baryonic dissipation because, without it, the
distribution of galactic halos and cluster halos is much flatter than the ZOE.
To explore the slope more carefully, we attempt to calculate the offset of
DHGs from E-rich groups and clusters in Figure~12a. Since both of these groups
are on the border of the ZOE, understanding their offset from one another will
fix its slope.

The calculated offset consists of three parts, which we treat as vectors. The
first vector, $V_1$, is the dissipation vector just calculated. The second
vector, $V_2$, is the slope of the density fluctuation spectrum between the
mass of a typical DHG dark halo and an E-rich cluster, assuming that both
types of object arise from the same $n$-$\sigma$ fluctuation.  The shift along
this vector assumes no change in $M/L$.  The third vector, $V_3$, takes
account of the fact that there is an extra diminuation in the surface
brightness of clusters relative to galaxy halos owing to the fact that much of
the baryons in clusters are in hot gas that never formed into galaxies.

To calculate $V_2$, we need the density fluctuation spectrum over the mass
range from DHG halos to E-rich clusters ($10^{12.5} - 10^{15}$ M$\odot$). We
take a power-law spectrum of the form

\begin{equation}
\delta_{rms} \propto M^{-1/2 -n/6},
\end{equation}

\noindent where $\delta_{rms}$ is the rms amplitude of perturbations of mass
$M$ at fixed, initial redshift.  We choose the value $n = -1.8$, which fits
the slopes of current candidate density fluctuation spectra over this mass
range (e.g., CDM; BFPR).  This yields $\delta_{rms} \propto M^{-0.2}$. 
After collapse, the equilibrium structural properties of a dark halo that began
with given $\delta_{rms}$ and $M$ obey the relations (Gott \& Rees 1975):

\begin{equation}
r \propto {\delta_{rms}}^{-1}~ M^{1/3},
\end{equation}

\begin{equation}
\rho \propto {\delta_{rms}}^{3},
\end{equation}

\noindent
and

\begin{equation}
V \propto {\delta_{rms}}^{1/2}~ M^{1/3}.
\end{equation}

\noindent
Combining these with $\delta_{rms} \propto M^{-0.2}$ yields the scaling laws 
for a typical object versus mass:

\begin{equation}
r \propto M^{0.53},
\end{equation}

\begin{equation}
\rho \propto M^{-0.60},
\end{equation}

\begin{equation}
V \propto M^{0.23},
\end{equation}

\noindent
and

\begin{equation}
I \propto M^{-0.07}.
\end{equation}

\noindent
assuming no change in $M/L$.  These relations allow us to calculate the vector 
$V_2$ assuming $n = -1.8$.

The final vector $V_3$ requires us to estimate the extra diminuation in
surface brightness of clusters relative to galaxy halos due to hot gas. We
estimate this empirically by taking the total change in $M/L$ between an
average DHG galaxy and a typical E-rich cluster (taken to be 0.95 dex in 
$\kappa_3$, or a factor of 44 in $M/L$, cf. Figure~11.).  We then 
divide this by the factor due to dissipation (10, above), which leaves 4.4 due
to hot gas. This is not far from the factor of 3 estimated from X-rays by
Muzhotsky (1991). Vector $V_3$ affects surface brightness only, leaving mass 
unchanged.

The three vectors are shown in Figure~12a.  Their components in $\kappa$-space
units are:  $V_1$ = (0.71, -2.04), $V_2$ = (1.79, -0.11), and $V_3$ = (0.00,
-1.05). Their sum is a little steeper than the observed ZOE, i.e., the surface
brightnesses of E-rich clusters are a little higher than expected.  This could
reflect the fact that dissipation is a little smaller than the factor of 10
assumed, or that visible galaxies in E-rich clusters have condensed slightly
relative to the DM by dynamical friction.  Given the roughness of the
calculation, the agreement is acceptable.

In closing this overview of structural properties, for completeness we plot
three other parameter combinations that are often useful. Figures~13 and 14
are updates of plots in BFPR. The first plots baryon density from Column 19
versus virial temperture from Column 18 in Table~1, using the equations given
in Appendix~\ref{app-1}. This combination is useful for assessing baryon
cooling and dissipation, and the density axis (assuming similar collapse
factors) is a measure of the redshift of collapse. The distribution of groups
and clusters in this diagram is similar to that in BFPR, but the galaxies are
different. The present distribution of galaxies is much more horizontal than 
the schematic, elongated representation in BFPR, with galaxies strung out at 
roughly constant density from right to left as a function of Hubble type.

The surprising conclusion from this graph is the roughly {\it similar
density} of all Hubble types within $r_e$ today.  At face value, this implies
similar collapse redshifts, an inference that could break down, however, if
all galaxies did not have constant collapse factor $f_{bary}$ (see below). Our
use of stellar half-light radii also underestimates the baryonic radii of
late-type galaxies, whose baryons are mostly H~I.  A switch to H~I radii would
move these galaxies to lower densities and later formation times.  Finally,
the present database is strongly biased against low-density, low-surface
brightness galaxies, which also would appear higher (later) in the diagram.
For all these reasons the apparent constancy of formation redshift for all
Hubble types in Figure~13 may be exaggerated. Nevertheless, the present data
are adequate to lilluminate the early Hubble types, and for them the flatness
of the locus in Figure~13 is striking.

Figure~14 plots velocity versus mass.  This figure resembles the corresponding
figure in BFPR more closely because this projection is a nearly edge-on view
of the metaplane, and the distribution of points is therefore insensitive to
the distribution of objects in the plane.  The long loci of galaxies here are
basically the Faber-Jackson and TF relations.  There is clearly a similar
relation for groups and clusters although it has never before received much 
notice.

Figure~15 is new here and plots radius versus mass. It is useful for
estimating the radial collapse factor for galaxies due to baryonic
dissipation. If we fit an $n = -1.8$ power-law slope through the middle of the
spiral-rich groups, compare to a similar locus through the middle of the
spirals themselves, and demand that the shrinkage due to dissipation be the
same in mass as in radius (i.e., both equal to $f_{bary}$), we find that an
offset of a factor of 19 in radius and mass is required to account for the
offset of spirals from spiral-rich groups in Figure~15. This value of
$f_{bary}$ is larger than the usual value of 10 assumed above, but the method
is rough. The point is that Figure~15 provides another target to shoot at in
matching the radii of galaxies versus groups.

\subsection{Hubble Types and the Dressler Effect}
\label{kaphcm-2}

We turn now to substructure within the galaxy and cluster subregions. The most
obvious feature is the ``banding" in $\kappa_1$--$\kappa_2$ for galaxies,
yielding the downward march of Hubble types away from the ZOE (Figure~6). A
natural (but not unique) interpretation of this phenomenon was given by Faber
(1982a) and BFPR in terms of forming different Hubble types from differing
degrees of overdensity. The overdensity within a given sphere of mass $M$ can
be written

\begin{equation}
\delta = n~ \delta_{rms}(M),
\end{equation}

\noindent where $\delta_{rms}(M)$ is the rms perturbation in spheres of mass
$M$.  Such a perturbation obeying Equation~28 is termed an ``$n$-$\sigma$''
perturbation.  Equations~21--23 imply that families of perturbations with
constant values of $n$ form parallel lines in $\kappa_1$--$\kappa_2$,
with higher $n$ lying at higher $\kappa_2$.  The relation between $n$ and 
the vertical offset $\delta_{2:1}$ is:

\begin{equation}
\Delta \log~n = 0.41~ \Delta \delta_{2:1}.
\end{equation}

\begin{deluxetable}{cccc}
\label{Nsig}
\tablenum{2}
\tablewidth{0pt}
\tablecaption{$n$-$\sigma$ Overdensities of Hubble Types}
\tablehead{\colhead{Type} & \colhead{$\Delta \delta_{2:1}$} &
\colhead{$n$-$\sigma$ (gE $\equiv$ 3)} &
\colhead{$n$-$\sigma$ (Sc $\equiv$ 1)}}
\startdata
gE     &  0.00 & 3    & 1.81  \nl
Sa     & -0.20 & 2.48 & 1.50  \nl
Sab,b  & -0.21 & 2.46 & 1.48  \nl
Sbc    & -0.36 & 2.14 & 1.29  \nl
Sc     & -0.63 & 1.66 & 1     \nl
Scd    & -1.34 & 0.85 & 0.51  \nl
Sd     & -1.97 & 0.47 & 0.28  \nl
Irr I  & -2.46 & 0.30 & 0.18  
\enddata
\end{deluxetable}

We can use this relation to infer relative values of overdensity $n$ for each
of the Hubble type bins in Figure~6.  These values are tabulated in Table~2.
Two normalizations have been used to convert to absolute $n$, one assuming
that gEs are 3-$\sigma$ perturbations as in BFPR, the other that Sc's are
1-$\sigma$ perturbations.  The latter is more plausible since the predicted
number of early-type galaxies is otherwise smaller than observed.  Either way,
the total range in $n$ implied by $\delta_{2:1}$ by Equation~29 is about a 
factor of 10.  Regardless of normalization, the implied overdensity for Irr's 
is only $0.2-0.3$, which is close to 0.  These objects barely managed to 
collapse.

\placetable{Nsig}

Notice that Equation~29 between $\delta_{2:1}$ and $n$ is logarithmic, which
means that, as $n$ approaches 0, the ridgelines in Figure~12a march to
$-\infty$ in $\kappa_2$.  The distribution of objects formed from an ensemble
of Gaussian random perturbations will therefore have a skewed distribution in
$\kappa_1$--$\kappa_2$; the density of objects peaks around the 1-$\sigma$
ridgeline and falls off above and below that due to the scarcity of
high-$\sigma$ peaks on the one hand and the spreading of low-$\sigma$ peaks to
large radii and low surface brightness on the other.  These features are not
yet quantifiable with the present data because they are not volume limited.
However, the observed distribution does broadly show the sharp upper envelope
and tail to low $\delta_{2:1}$ that are expected.

The logarithmic relation between $n$ and $\delta_{2:1}$ in Equation~29 may
explain the variable widths of the distributions observed for different Hubble
types in $\delta_{2:1}$.  If each Hubble type were selected from approximately
equal intervals in $n$, it would appear in $\kappa_1$--$\kappa_2$ with
progressively wider widths for lower values of $n$.  This may help to account
for the relatively tight loci of early Hubble types versus the wide, amorphous
loci of the later types.

The above discussion implicitly assumes all galaxies have the same radial
baryonic collapse factor, $f_{bary}$ and overdensity $n$ is the sole
determiner of Hubble type.  This may be termed the ``density hypothesis''
(Faber 1982a). An alternate picture is $f_{bary}$ varies with Hubble type. The
simplest way this might happen is if the dark-halo angular momentum parameter
$\lambda \equiv L E^{1/2} G^{-1} M^{-5/2}$ (Peebles 1969) is not constant with
Hubble type. If baryonic collapse is halted by angular momentum (Fall \&
Efstathiou 1980), more slowly rotating halos will collapse further.  If
$\lambda$ alone were the determining factor for Hubble type, early-type
galaxies would descend from slowly rotating, low-$\lambda$ dark halos because
they are denser at a given mass than late-type galaxies, while late-type
galaxies would come from rapidly rotating halos.  The dissipational collapse
factor $f_{bary}$ would then determine Hubble type.  Faber (1982a) termed this
the ``dissipation hypothesis'', but a better term might be the
``$\lambda$-hypothesis.''

Since the density hypothesis fits many features of the data, there is no
compelling reason at the present time to invoke $\lambda$ variations. N-body
simulations of dark halos have furthermore failed to reveal any strong
correlation between $\lambda$ and environment, as would be needed to explain
the high density of early-type galaxies in clusters (Barnes \& Efstathiou
1987).  However, {\it very} low-surface-brightness objects are
under-represented in the present data set. If they were taken into account,
the total spread of galaxies in $\delta_{2:1}$ would be larger, and there
would be more reason to invoke $\lambda$, at least as a second parameter.
Higher $\lambda$ in later-type galaxies would cause them to collapse less,
leading to more DM within their baryonic radii (Tinsley 1981, Faber 1982a,
Verheyen 1997). Excess DM would tend to suppress spiral structure (Toomre
1964, 1981; Binney \& Tremaine 1987) in later Hubble types, producing Irr's as
observed.  On balance, it would seem timely to re-examine the $\lambda$--
versus density-- hypotheses using the newer simulations that model baryonic
dissipation more accurately. It is of course possible that both $\lambda$ and
initial overdensity are play a role in determining Hubble type.

Turning now to groups and clusters, we note a parallel banding effect that
matches the Hubble type bands among galaxies.  The analog to Hubble type is
mean group Hubble type and, as previously noted, there is a strong tendency for
early-type-dominated groups to lie close to the ZOE and for groups to become
progressively late-type-dominated away from this line.  This is simply the
``Dressler effect" (Dressler 1980, Postman \& Geller 1984), wherein the Hubble
type of a galaxy is strongly correlated with its environment --- early types
are found preferentially in dense regions, while late types are found in
sparse regions. The origin of this effect is probably that realistic density
fluctuation spectra are flatter than white noise, with the result that density
fluctuations on neighboring scales are correlated (BFPR; Bardeen, Bond, White,
\& Efstathiou 1988). The high-$\sigma$ peaks that spawn early-type galaxies
(in the density hypothesis) are statistically embedded in high-peak clusters
(BFPR), which also collapse to high density and lie near the ZOE. This effect
has been born out by N-body/hydro simulations (e.g., Cen \& Ostriker 1994).

\subsection{The Slope of Hubble Types within $\kappa_1$--$\kappa_2$}
\label{kaphcm-3}

So far the HCM theory of structure formation is adequate to account for most
of the broad features of structures in $\kappa$-space. We now highlight one
aspect of the data that has so far eluded explanation.  The question at issue
is the {\it slope} of the galaxy loci for individual Hubble types in
$\kappa_1$--$\kappa_2$. We have seen in Figures~1--6 that this slope follows
the ZOE. We were able to account for the slope of this line {\it between
galaxies and clusters} by assuming an $n = -1.8$ power-law coupled with
plausible dissipation.  However, the simplest theory would say that
dissipation and overdensity ($n$) are {\it both constant within a given Hubble
type}.  Each type should therefore parallel the density fluctuation slope
(vector $V_2$ in Figure~12b). This is much shallower than the observed loci
for Hubble types.

Let us review exactly what the vector $V_2$ means.  It is the predicted slope
for a collection of objects formed from a power-law initial density fluctuation
spectrum with slope $n = -1.8$, constant $n$-$\sigma$ overdensity, constant
$M/L$, and constant radial baryonic collapse factor, $f_{bary}$.  Since the
observed slope does not match this prediction, one or more of these
assumptions must be wrong.  $M/L$ is constrained to vary only modestly by
observations, and the constant slope and overdensity of a given Hubble type
are attractive assumptions we are wanting to test.  That leaves $f_{bary} = $
constant, which is perhaps the least secure assumption. In words, the radii of
big galaxies within each Hubble type seem to be larger than predicted by the
constant collapse picture --- big galaxies are more diffuse than expected.

This problem was first highlighted for DHGs by B$^2$F1 but now appears to
affect {\it all} Hubble types through at least Sc. What is needed in the
constant collapse picture is to either halt the collapse of the baryons at
larger radii in the first place or reinflate them to larger radii later.  The
extension of the problem to disk-dominated galaxies is thus very significant.
Pressure-supported DHGs can be inflated by simply adding energy, for example
by heating their stars through mergers in dense groups.  However, increasing
the radii of {\it disk} galaxies is not as easy --- it requires adding angular
momentum as well as energy.  This cannot be done through mergers without
destroying the disks. Thus, the process, whatever it is, must take place for
disk galaxies while the material is still gaseous, which adds an important
constraint.

To summarize, understanding the existence of the ZOE involves two separate
phenomena.  One issue is the slope of the line between galaxies and clusters,
and that seems explicable using HCM together with plausible dissipation.  The
other is the slope of individual Hubble types themselves. This problem has no
easy solution at the present time.  Is the existence of one, smooth ZOE an
accident, or is it fundamental?  Whatever the answer, matching galactic radii
as a function of mass and Hubble type is one of the most pressing targets for
galaxy simulations.

\subsection{Mass-to-Light Ratios of Groups and Clusters} 
\label{kaphcm-4}

A final puzzle concerns the pronounced increase (and curvature) in $\kappa_3$
vs. $\kappa_1$ ($M/L$ vs. $M$) for groups and clusters (see Figure~9). An
offset of groups and clusters to higher $M/L$ than galaxies is expected
because the latter contain more DM.  However, this offset does not appear
suddenly in going from galaxies to groups, as one might have expected, but
rather sets in as a gradual rise in $M/L$ vs. group size.  

The Cold + Hot Dark Matter (CHDM) simulations of Nolthenius, Klypin, \& 
Primack (1997; hereafter NKP97) may help to explain part of this effect.  The
apparent overlap in $M/L$ between small groups and galaxies may be caused the
large errors of small groups due to small-number statistics and projection
effects. Appendix~\ref{app-2} compares $M/L$'s for redshift-selected small
groups vs. the same groups selected in real space (Figures~16a and 16b).  The
lower $M/L$ values of the former overlap with the $M/L$'s of galaxies, while the
latter do not.  The locus for the simulation redshift-selected groups is also
steeper at the faint end and follows the mass-error trajectory.  The same
effect is seen in the real groups. The simlulations thus suggest that errors
may be obscuring a real discontinuity in $M/L$ between galaxies and small
groups and creating the appearance of a smooth trend.  On the other hand, the
fact that Malin 2 lies on the gE FP, while Malin 1 lies within the group $M/L$
for its $M$ suggest that these very rare kinds of galaxies can bridge this gap.

This effect cannot produce a second and more important trend --- the striking
overall {\it rise} in group $M/L$ as a function of mass.  Remarkably, the
same effect is seen in the simulations as well as the real data (Figure~9 vs.
Figures~16a and 16b).  Tests show that the effect persists for ``break-up''
vs. ``no-break-up'' group catalogs, and for real-space vs. redshift-space
selected groups. It is not due to a decrease in the average luminosities of
galaxies in large groups (this effect is 20-30\% at most, Appendix~\ref{app-2})
but rather to an actual decrease in the number of galaxies identified in large
clusters.  As we emphasize in that Appendix, the simple prescription used for 
identifying and illuminating galaxies in these pure dissipationless simulations
should be adequate only for comparing the properties of groups selected in 
redshift space to their counterparts selected in real space. Even so, the 
match to the N93 groups is good enough that it's tempting to wonder whether 
the simulations may yet capture more truth than one might at first suppose.
In addition, it remains unresolved why the slope of $M/L$ for large groups so
closely matches that for DHGs (the two planes are parallel, cf. Figure~10).  
All-in-all, there are several important unanswered questions concerning the 
$M/L$'s and galaxy formation efficiency in groups and clusters.

\subsection{Guidelines for Future Comparisons to Simulations} 
\label{kaphcm-5}

It is our hope that the present database (and its descendents) will be of use
to modelers attempting to simulate galaxy formation and other types of
structure in the universe. In assembling the present data, we have taken the
easy path of sticking close to raw observations with a minimum of
transformations to more fundamental quantities.  This has the virtue of
minimizing present uncertainties but forces future simulations to be more
complete and realistic. Here is a reminder of the basic properties of the
present data:

1) All luminosities are B-band light emitted by stars.  

2) The effect of interstellar extinction in our Galaxy is fully corrected for,
but the extinction correction for internal extinction corrects galaxies only
to face-on. The correction from face-on to fully-corrected is not well
understood and depends on both the ratio of absorption to scattering for dust
grains and their distribution in the galaxy relative to the stars.  If forward
scattering by dust is dominant, face-on luminosities may even be too bright.

3) Radii are half light radii defined by starlight in the observed B-band 
light profile.  No adjustments are made for variable extinction with radius.

4) The central line-of-sight velocity dispersion for a hot system, $\sigma_c$,
is adopted as the fundamental definition of velocity.  Effective dispersions,
$\sigma_e$, of groups and clusters and peak rotation velocities, $V_{rot}$, of 
disk systems are transformed to equivalent values of $\sigma_c$.  

5) The basic finding list for galaxies consists of objects with the necessary
data in the RC3.  The sample is therefore quasi-magnitude limited and weighted
to bright objects.  Galaxies of low surface brightness are especially selected
against.

Four physical processes must be modeled to permit comparison with this data 
set: 

1) Stellar radii of galaxies must be determined.  This requires a theory for
the initial distribution of gas within the galaxy, plus a theory for its
conversion to stars as a function of radius and time. The gaseous radii of
disks depend on the fraction of baryons that have cooled by today, and from
what radius they fell in (the collapse factor). This in turn depends on
angular momentum generation in dark halos including the exchange of angular
momentum between the baryons and DM and among the various baryonic components.
Angular momentum transfer in mergers and in distant interactions also needs
to be considered.

The stellar radii of DHGs depend on the radii of any previously formed stellar 
disks absorbed in mergers (Kauffmann, White, \& Guiderdoni 1993), the relative 
timescales for infall versus star formation (Faber 1982b), mass loss in 
stellar winds, merger dynamics, and possibly other unknown factors.

2) B-band luminosities of galaxies must be computed.  This requires the
history of star formation at every location, including the stellar birthrate
and initial mass function.  Luminosity profiles and total luminosities need to
be converted to observed face-on quantities by correcting for dust.

3) For disk galaxies, observed peak rotation velocity $V_{rot}$ must be related
to $\sigma_e$ for the dark halo.  This requires a theory for gas infall and 
the quasi-adiabatic simultaneous pulling-in of the DM particles.

4) For groups and clusters, the main uncertainty is what fraction of 
baryons is converted into stars and how this varies as a function of local 
galaxy density.  There may also be radial or other segregation effects between
galaxies, hot gas, and DM.  

\section{Conclusion}
\label{kapconcl}

In this paper we have extended our view of $\kappa$-space to include all major
types of equilibrium, self-gravitating systems in the local universe.  To do
this, we have developed consistent definitions of luminosity, radius, surface
brightness, and internal velocity.  Internal velocities in particular require
special transformations to achieve uniformity among the different types of
galaxies. When these corrections are applied, a pleasing continuum of physical
properties is evident among galaxies of all Hubble types from giant 
ellipticals to Sd and Irr galaxies.

Each type of stellar system is found to populate its own fundamental plane in
$\kappa$-space. Six different planes are found: 1) the original fundamental
plane for DHGs; 2) a parallel plane slightly offset for Sa-Sc spirals; 3) a
plane with different tilt but similar zero point for Scd-Irr galaxies; 4) a
plane parallel to the DHG plane but offset by a factor of 10 in mass-to-light
ratio for rich galaxy clusters; 5) a plane for galaxy groups that smoothly
bridges the gap between rich clusters and galaxies; and 6) a plane for
Galactic globular clusters.  We propose the term ``cosmic metaplane'' to
describe this ensemble of interrelated and interconnected fundamental planes.

The projection $\kappa_1$--$\kappa_3$ ($M/L$ vs. $M$) views all planes nearly
edge-on. The $\kappa_1$--$\kappa_2$ projection views all planes close to
face-on, while $\kappa_2$--$\kappa_3$ shows variable slopes for different
groups owing to the different tilts of the individual planes. No stellar
system yet violates the rule first found from the study of DHGs, namely,
$\kappa_1 + \kappa_2 < 8$, which we term the ``zone of exclusion,'' or ZOE.  
In physical terms, this implies that the maximum global luminosity volume 
density of stellar systems varies as $M^{-4/3}$.

Each Hubble type defines a band in $\kappa_1$--$\kappa_2$ parallel to the ZOE.
Hubble types march monotonically away from the ZOE, with DHGs being closest
and Sd--Irr's most distant.  This reflects a decrease in the average mass and
surface brightness along the Hubble sequence. The Tully-Fisher relation ($L
\propto V_{rot}^{3-4}$) is simply the proper compromise projection to view the
spiral-irregular planes close to edge on, analogous to the $D_n$-$\sigma$
relation for DHGs.  All galaxy types, spirals/irregulars and DHGs, show the
mutual dependence of characteristic velocity, effective surface brightness and
effective radius that first characterized the fundamental plane (FP) for gE
galaxies.  The fact that the coefficients in these relationships differ
somewhat among galaxies of different Hubble types again illustrates the fact
that we are not dealing with just one FP, but several interlocking ones that
form a cosmic metaplane.

The rich clusters orignally studied by Schaeffer et al. (1993) define yet
another fundamental plane that is remarkably tight and parallel to that of the
DHGs but offset by a factor of 10 higher in mass-to-light ratio. The galaxy
groups of Nolthenius (1993) likewise define a fundamental plane, but one that
is tilted and appears to link the FPs for rich clusters and for DHGs.  Some
fraction of this tilt is due to errors in group properties due to small number
statistics and chance projection effects, but most appears to be real. Just as
the loci of Hubble types march steadily away from the ZOE along the Hubble
sequence, the mean location of groups marches away from the ZOE with
increasing spiral content.   This is the well known Dressler (1980) effect.
The two large diffuse galaxies known as Malin 1 and Malin 2 have physical
properties similar to those of spiral-dominated galaxy groups.

Globular clusters distribute themselves within $\kappa$-space close to the DHG
fundamental plane (cf. Djorgovski 1995; Schaeffer et al. 1993) but offset in
zeropoint.  They occupy a small band of radii, as predicted if the present
population of clusters were a small surviving subset of a larger, broader
population.

In broad-brush strokes, the distribution of galaxies, galaxy groups and galaxy
clusters within $\kappa$-space can be understood as the result of hierarchical
clustering and merging from a power-law initial density fluctuation spectrum.
The cosmic metaplane is simply the cosmic virial plane common to all
self-gravitating equilibrium stellar systems, tilted and displaced in
mass-to-light ratio by different amounts in different regions due to
differences in stellar populations and amount of baryonic dissipation.
Hierarchical clustering from an $n = -1.8$ power-law density fluctuation
spectrum (plus dissipation) comes close to reproducing the slope of the 
ZOE.  The progressive displacement of Hubble types from this
line is consistent with the formation of early-type galaxies from higher
$n$-$\sigma$ fluctuations than later Hubble types.

A major mystery is why the slopes of {\it individual} Hubble types in the
$\kappa_1$--$\kappa_2$ plane parallel the ZOE.  At face value, this appears to
suggest {\it less} dissipation of massive galaxies within their dark halos
compared to lower-mass galaxies of the same Hubble type.  A second mystery is
the behavior of $M/L$ vs. $L$ for groups and clusters. Two slopes are seen,
both tight, with group $M/L$'s climbing more steeply than clusters. The same
general rise in $M/L$ vs. $M$ is seen in the simulations, but we have no good
physical explanation either there or in the real data.  The tightness in $M/L$
implies a remarkably close coupling between the galaxy formation efficiency,
stellar populations, hot gas content, and dark matter content of groups and
clusters as a function of total mass.

We discuss the limitations of the present data and alert modelers to the large
number of physical processes that need to be calculated in order to make a
realistic comparison between these data and simulations. Use of the B band is
an important limitation. One would ideally like to view the properties of
these stellar systems in redder passbands less affected by dust and recent
star formation. Also crucial is a volume-limited sample of nearby galaxies,
needed for an unbiased census of galaxies in $\kappa$-space.

We now realize that the initial discovery of the elliptical galaxy fundamental 
plane a decade ago was but the first glimpse of the full cosmic metaplane that 
is defined by the physical properties of all gravitationally-bound, star-defined
systems.  As with our first investigation of dynamically hot galaxies,
$\kappa$--space has proved to be a valuable tool for exploring this
``cosmic metaplane.''  

\acknowledgments

The data used here are the result of the collected efforts of many
astronomers over the years, as summarized in the catalog tables we have used.
We thank Alberto Cappi for sending us the Schaefer et al. data in computer
readable format, Giampaolo Piotto for pointing us the in right direction
for the globular cluster data, and Anatoly Klypin and Joel Primack for use of 
their CHDM simulation.  DB acknowledges partial support from NSF AST90-16930.  
SMF acknowledges support from NSF AST95-29008 and from NAS-5-1661 to the
WFPC1 IDT.

\newpage

\appendix

\section{Standard Quantities Derived from $\kappa$ Parameters}
\label{app-1}

In this Appendix we give convenient relations for transforming $\kappa$ 
parameters into more familiar quantities. The fundamental plane parameters 
$r_e$, $I_e$ and $\sigma_c$ are related to the $\kappa$ parameters as follows:

\begin{equation}
\log r_e = 
   \frac{\sqrt{2}}{2} \kappa_1 
  -\frac{\sqrt{6}}{6} \kappa_2 
  -\frac{\sqrt{3}}{3} \kappa_3 \, ;
\end{equation}

\begin{equation}
\log I_e = 
   \frac{\sqrt{6}}{3} \kappa_2 
  -\frac{\sqrt{3}}{3} \kappa_3 \, ;
\end{equation}

\begin{equation}
\log \sigma_c = 
   \frac{\sqrt{2}}{4} \kappa_1 
  +\frac{\sqrt{6}}{12} \kappa_2 
  +\frac{\sqrt{3}}{6} \kappa_3 \, .
\end{equation}

\noindent With $r_e$ in kpc and $\sigma_c$ in km s$^{-1}$, we define 
$M_e = 4.65 \times 10^5 \sigma_c^2 r_e$ M$_\odot$, where we have taken the 
standard Keplerian formula for defining the mass of a disk, 
$M_e = r_e V_{rot}^2/G$, and corrected $V_{rot}$ to $\sigma_c$ as given by 
$K_2$.  This yields

\begin{equation}
\log M_e \, ({\rm M}_\odot) 
   = \log \, (\sigma_c^2 r_e) + 5.67 \, 
   = \sqrt{2} \, \kappa_1 + 5.67 \, .
\end{equation}

\noindent With $r_e$ in kpc and $I_e$ in B-band 
L$_\odot$ pc$^{-2}$, the luminosity 
within the effective radius is defined as $L_e = \pi \times 10^6 \, I_e 
{r_e}^2$ L$_\odot$, resulting in

\begin{equation}
\log L_e \, ({\rm L}_\odot) \, 
  = \log \, (I_e r_e^2) + 6.50 \, 
  = \sqrt{2} \kappa_1  - \sqrt{3} \kappa_3 + 6.50 \, .
\end{equation}

\noindent The B-band mass-to-light ratio within $r_e$ is
the ratio of these two: 

\begin{equation}
\log \, [M_e/L_e] \, (\odot) 
   = \log \, (\sigma_c^2 r_e) - \log \, (I_e {r_e}^2) - 0.83 \, 
   = \sqrt{3} \kappa_3 - 0.83 \, .
\end{equation}

Our definitions of effective virial temperature, $T_e$, and mean baryon number
density within the effective radius, $n_{bary,e}$, follow those of BFPR. For
$T_e$ we equate a Maxwellian distribution to a Boltzmann distribution, $3/2
\mu m_p \sigma_e^2 = 3/2 k T$, where $m_p$ is the proton mass.  Using a mean
molecular weight $\mu = 0.6$ for ionized primordial H$+$He yields $T_e =
72.7~\sigma_e^2$ K $= 51.4~\sigma_c^2$ K, for $\sigma$ in km s$^{-1}$.  Thus

\begin{equation}
\log T_e \, ({\rm K}) 
   = \log \, (\sigma_c^2) + 1.71
   = \frac{\sqrt{2}}{2} \kappa_1 
   +\frac{\sqrt{6}}{6} \kappa_2 
   +\frac{\sqrt{3}}{3} \kappa_3 + 1.71 \, .  
\end{equation}

The baryon number density $n_{bary,e}$ is defined as $f_{bary} M_e/(4/3~\pi
r_e^{3} m_p)^{-1}$, where $f_{bary}$ is the fraction of $M_e$ contributed by
baryons within the effective radius.  The baryon correction is expressed as
$s_{type} = {\rm log}~f_{bary}$. In the present units,

\begin{equation}
\log n_{bary,e} \, ({\rm cm}^{-3}) 
   = \log \, (\sigma_c/r_e)^2 - 2.34 + s_{type}    
   = -\frac{\sqrt{2}}{2} \kappa_1 
   +\frac{\sqrt{6}}{2} \kappa_2 
   +\sqrt{3} \kappa_3 - 2.34 + s_{type} \, .
\end{equation}

Systems that are dominated by dark matter (DM) have significant baryon
corrections. Examples are galaxy groups (Mulchaey et al. 1996), galaxy clusters
(Mushotzsky 1991), and extreme dwarf spheroidals (B$^2$F1), for which we
assume $s_{type} = -1.0$.  For dwarf ellipticals, we assume $s_{type} = -0.2$
(B$^2$F1), and for spirals we also assume $s_{type} = -0.2$ dex (Rubin et al.
1985, corrected to $r_e$). All of these $s_{type}$ corrections are likely to
be within 0.3 dex of being correct, which is similar to the error that we make
by assuming that all the mass is spherically distributed.  Systematic errors
of this size are of no consequence to our conclusions.

\section{Using N-Body Simulations to Calibrate Group $\kappa$ Properties}
\label{app-2}

There are several well-known difficulties in linking galaxies in redshift 
space into true, physical associations with properly measured properties:

(1) As filamentary structures are usually comprised of a series of dense
regions of galaxies, groups are often poorly separated from surrounding
galaxies. This can lead to contamination from foreground and background
galaxies.

(2) Group structure is often unrelaxed, so that local gravitational tidal
fields from larger masses (say superclusters) can have a significant effect on
the spatial and redshift distribution of galaxies within even a bound group.
 
(3) Groups generally have few bright, visible members, resulting in shot
noise and random projection effects which can lead to substantial errors in
determined quantities.

These problems require that a grouping algorithm be carefully calibrated
against realistic simulations which include these effects.  While based on the
early Nolthenius \& White (1987) simulations, the N93 link criteria have
recently been shown to be near optimal (as defined below) by calibration with
more recent N-body simulations (Nolthenius, Klypin \& Primack 1994; NKP94;
Nolthenius, Klypin \& Primack 1997; NKP97), which use the simulations of
Klypin, Nolthenius \& Primack (1997). Our plan is to use versions of these
catalogs constructed to match the CfA1 Survey and compare $\kappa$--space
properties in redshift space to those of groups selected in real space (i.e.
using full 3D information).

\subsection{Constructing Mock-Sky Catalogs}
\label{app-2-1}

The Klypin et al. N-body simulation we use here evolves an $\Omega=1$,
$70\%/30\%$ mix of ``Cold + Hot Dark Matter'' (a CHDM model) on a 100 Mpc
across $512^3$ cell grid using a particle-mesh pure dissipationaless code.
Within this simulation, potential galaxies are identified as cells with local
overdensity $\delta \rho / \rho > 30$. Since galaxies must be local
overdensities, a centrally concentrated array of overdense cells will still be
identified as only a single galaxy. We identify all $\sim 30000$ ``galaxies'',
generate an equal number of Schechter luminosities, and pair up the
luminosities with galaxies so that luminosity rises monotonically with 1-cell
mass. 

We then select a ``home galaxy'' similar to our location in the real universe,
observe the catalog and impose the CfA1's $m_B=14.5$. magnitude limit. Since
``observing'' is done after $L$'s are assigned, large scale flows and
Malmquist bias will lead to a steeper $\alpha$ and brighter $M^*$ when
observed in redshift space.  The $L$ assignment is therefore done iteratively
until the redshift-space-observed luminosity function and the sky-projected
galaxy density of the resulting catalogs both match that of the CfA1. To
include possible edge effects yet maximize sample size, we include the full
sky minus a $20\deg$ wide ``zone of avoidance'', which includes roughly 9000
``galaxies'' and 830 ``groups''. See NKP97 for a more complete description of
the construction of the sky catalogs.

We analyze here NKP97's CHDM$_2$ simulation because both the fraction of
galaxies in groups and the median group velocity dispersions closely match
those of identically selected CfA1 groups not only at the optimal link
parameters but throughout the entire range of links (NKP94, NKP97).  
NKP97 argue that these two properties are very sensitive to
the spatial and velocity structure of galaxies on group scales, and therefore
are the most appropriate indicators to match when attempting to
design and calibrate an optimal grouping algorithm.

However, we strongly caution against identifying these objects with
representative observed galaxies {\it in a CHDM universe}, for two reasons.
First, with a cell size of 195 kpc, these simulations have neither the
resolution nor the baryonic physics needed to model individual galaxies or
even galaxy dark matter halos with confidence. Second, the dark matter halos
in these simulations are affected by the overmerging problem (Katz \& White
1993) and the proper way to ``break up'' these overmergers into galaxies is
still poorly understood.  In fact, it is the un-broken catalogs which show the
closest structural similarities to the real data and which we therefore use
here. We also note that while the NKP97 ``breakup'' catalogs yield closely
similar $\kappa$--space properties, it is quite possible the level of
overmerging was underestimated (see NKP97 ``2$\nu$'' results). Our goal here
is only to see how $\kappa$ values change between real space and redshift
space. We resist the temptation to judge the merits of CHDM as a cosmological
model.

The redshift space link relations are described in N93 and NKP97. To select
corresponding groups in real space we again seek to identify number density
enhancements above a consistent density contrast, With 3D information, the N93
sky and redshift dimensions are now described by a single isotropic
separation, which is again scaled to account for growing incompleteness at
higher distance. Distances here assume a smooth Hubble law, and, unlike in
N93, there has been no attempt to ``fine-tune'' groups in the densest regions.
Thus, mock-galaxies at distance $V$ from the observer are linked if their 3D
separation is less than $D(V)_L$, defined by

\begin{equation}
\rm \lambda(V) = {L(V)_{lim}\over{L^*}} \quad ,
\end{equation}

\begin{equation}
\rm R(V)_{\rm M.I.S.} = {\left[ \Phi(V) \right]}^{-1/3} = {\left( \phi^{*} \,
\Gamma \left[ 1+\alpha,\lambda(V) \right] \right)}^{-1/3} \quad ,  
\end{equation}

\centerline{and}

\begin{equation}
\rm D(V)_L = D_0 \, R(V_0)_{\rm M.I.S.} \, {\left[ \Phi(V_0) \over 
\Phi(V) \right]}^{1/3} .  
\end{equation}

\noindent $\Phi$ is the integrated galaxy luminosity function above the
limiting luminosity $\rm L(V)_{\rm lim}$ visible at distance V due to the
catalog apparent magnitude limit $\rm m_{\rm lim}$ (Schechter 1976) with the
usual parameters $\phi^*$ and $\alpha$; $\rm L^*$ is the luminosity
corresponding to the Schechter luminosity $\rm M^*$; $\rm \lambda(V)$ is
defined as $\rm 10^{[0.4(M^*-(m_{\rm lim} - 25 - 5 \log (V/H_0)]}$; $H_0$
is the Hubble constant; $\Gamma$ is the incomplete gamma function, $\rm D_0 =
0.36$ is our link parameter, $\rm V_0$ is an arbitrary scaling distance set to
1000 km sec$^{-1}$, and $\rm R(V)_{\rm M.I.S.}$ is the mean intergalaxy
spacing. The $\kappa_{\rm 3D}$ parameters are calculated using one-dimensional
velocity dispersions obtained by scaling the true 3D velocity dispersions
assuming isotropy (i.e. no strongly radial/tangential orbits for galaxies in
mock groups). We define the $\kappa$ parameters as before, except that we use
full 3D information to define the half light, virial radius, and velocity
dispersion. The distribution of the resulting groups in $\kappa$--space is
shown in Figure~16b.

The optimal linking in redshift space is defined as that which produces median
group velocity dispersions and group memberships which best match those formed
from grouping in real space in the mock-sky catalog. This optimal redshift
linkage, $V_5 = 350$ km/sec for $D_0 = 0.36$, was determined in NKP94 and
fortuitously agrees with that used earlier on the CfA1 data to generate the
N93 groups catalog used here.  The virial-to-true mass ratios of simulated
groups selected for the mock-sky catalog with the N93 link criteria have large
scatter, but show a median quite close to 1 (NKP97).  However, $\sim20-25\%$
of the simulated 3-member groups have such large virial-to-true mass ratios
that they are likely chance groupings.  This suggests that most CfA1 groups
are likely bound, but with some contamination from unbound groups at the
3-member level.  There is also confirming evidence from X-ray studies of
galaxy groups that at least the E-dominated groups are bound systems, even
with as few as 2 giant galaxies (cf. Mulchaey et al. 1996).

As in all of our plots, selection effects will skew the relative density of
groups across $\kappa$--space. In particular, we preferentially select richer
groups both because they can be seen at greater distance and because massive
potentials with few galaxies will tend to be missed since their large velocity
dispersions will mimic low density in redshift space.

\subsection{Results}
\label{app-2-2}

Figure~16a shows the $\kappa$--space distribution for these CHDM$_2$ groups
identified in redshift space using the N93 linking. The distribution of the 3D
selected simulation groups in $\kappa$--space is shown in Figure~16b.
Comparing Figure~16a with Figure~16b, we find that properties change little in
going from real to to redshift space, so that the fundamental features of the
kappa space distributions of observed groups, which of necessity must be
identified in redshift space, should be reliable.  We find shifts, in the
sense redshift-space to real-space, of only 0.3 in $\kappa_1$, -0.2 in 
$\kappa_2$, and 0.3 in $\kappa_3$. There is one important difference between 
redshift-selected
groups and with real-space selected groups, however. As the simulated volume
is small there is a paucity of rich dense clusters in the simulation. In
comparing these figures, we've chosen not to focus on individual $\kappa$
shifts for each group, since not all groups have both redshift--space and
real-space counterparts.  Some groups break into two, or merge, or members
drop out, or new members appear.

We next note that, as in the CfA1 data, there is somewhat of a tendency for
the richer groups to show a flatter fundamental plane than poor groups. At
least some of this tendency is due to correlated noise. Note that the (noisy)
virial mass appears in the numerator of both $\kappa_1$ and $\kappa_3$. From
Equations 1 and 3, we see that perfectly random masses would yield a
``fundamental plane'' with slope $\sqrt(2/3) = 0.82$. Since this is steeper
than observed, it's not suprising that more poorly sampled and hence noisier
groups would show a steeper FP. However, the fact that the slope is little
changed for 3D groups shows that most of slope of their FP is real.

We also find a tendency for the grouping algorithm to miss some valid members
on the tail of the redshift distribution for distant massive groups, while
including spurious infalling outliers along the line of sight. This latter
effect has the tendency to artifically concentrate the group, and thus lower
the virial radius and mass. High velocity dispersion systems are more common
at high distance.  If too many members on the tail of the redshift
distribution miss the magnitude cut, they can break the redshift link and miss
inclusion. Indeed, the algorithm was tuned to best select average groups and
appears to systematically underestimate the dispersion of massive clusters.
This may lead to the cutoff in $M/L$ for rich clusters that is somewhat
sharper than for the poorer, lower velocity dispersion groups. Of the 5 Abell
clusters in N93 which have Struble \& Rood (1991) velocity dispersions, the
N93 dispersions average 30\% smaller ($\Delta \log \sigma=-0.167$).  Some part
of this effect may also arise from subclustering within clusters.

Finally, we note the striking similarity between the observed (Figure~9) and
simulated distributions of groups in $\kappa$-space. To some extent, this may
reflect the careful choice of simulation and mock-sky constructions. On the
other hand, is it possible that the distributions are primarily determined by
the gravitation of the dark matter and our simple luminosity assignment method
is not far off. If so, the simulations may capture most of the essense of the
problem and, if correcting for overmerging leaves $\kappa$--space relatively
unchanged, the agreement between CHDM and observations may then be
significant. In any case, we found that the population differences between
rich and poor simulation groups was only a minor factor; richer clusters have
galaxies with luminosities only $20-30\%$ below those of poorer clusters.

The main cause for the $M/L$ vs. $M$ trend is that there are simply fewer 
galaxies per unit mass in richer groups.  This is seen by calculating
$d \log N / d \log M$, where $N$ is the number of visible group members at 
an arbitrary, but constant distance, and $M$ is the group virial mass.
We find $d \log N / d \log M = 0.57$, 0.50 and 0.68 for the redshift space, 
real space, and breakup versions, respectively.  While a more appropriate 
break-up prescription may well produce more galaxies (NKP97), this 
hypothetical break-up scheme would have to be quite extreme to make 
$d \log N / d \log M = 1$.

\newpage

\centerline{FIGURE CAPTIONS}

\vspace*{3mm}

\noindent {\bf Figure 1} The $\kappa$ parameters for spiral galaxies in the
Virgo cluster, and elliptical galaxies in Virgo and in the Coma cluster,
plotted in a three-dimensional fold-out of $\kappa$-space. Symbols as given in
the legend, used only for this figure. Dark lines represent $\pm$30\% distance
errors.  The dashed line in the $\kappa_1$--$\kappa_2$ plane is the zone of
exclusion line (ZOE) discussed in the text, $\kappa_1 + \kappa_2 = 8$   The
solid line in the $\kappa_1$--$\kappa_3$ plane is the gE fundamental plane.
The dashed and solid lines in the $\kappa_2$--$\kappa_3$ are the projection of
the fundamental plane in this projection.  Note how, in absence of distance
errors, spiral galaxies and elliptical galaxies occupy similar regions in
$\kappa$-space.

\noindent {\bf Figure 2} $\kappa$ parameters for dynamically hot galaxies
(DHGs), plotted in the same 3D fold-out manner as in Figure~1.  The plotting
symbols here are chosen to highlight the positions of isotropic DHGs (open
squares and diamonds) versus anisotropic DHGs (closed squares and diamonds) in
$\kappa$-space, as well as to separate giant DHGs from dwarf DHGs.  These data
come from B$^2$F1. Small open squares are galaxies from the 7Samurai data set
(Faber et al. 1989) with no isotropy measurements.

\noindent {\bf Figure 3} $\kappa$ parameters for Sa+Sab galaxies (broad
crosses) and Sb galaxies (open hexagons) plotted in the same 3D fold-out
manner as in Figure~1.  These early-type spirals are now farther from the 
ZOE in the $\kappa_1$--$\kappa_2$ plane, and lie below the DHG
fundamental plane in $\kappa_1$--$\kappa_3$. The two galaxies that lie just on
the ZOE are NGC~669 and the well-known bulge-dominated Sa
galaxy NGC~4594.

\noindent {\bf Figure 4} $\kappa$ parameters for Sbc galaxies (closed
triangles) and Sc galaxies (open circles) plotted in the same 3D fold-out
manner as in Figure~1.  These later-type spirals have now marched even further
from the ZOE.

\noindent {\bf Figure 5} $\kappa$ parameters for Scd galaxies (open
pentagons), Sd galaxies (stars), Irregular galaxies (dark crosses) and the two
very large spiral galaxies Malin 1 and Malin 2 (large crosses), plotted in the
same 3D fold-out manner as in Figure~1. These very late-type spirals occupy
roughly the same region as the dwarf DHGs (dE's and dSph's), suggesting that
the latter may have originated from the former via gas loss.  Two of the
lowest mass Irr galaxies in this sample (A~22044-1 = DDO 210 and Leo A = DDO
69) lie intermediate in $\kappa$ properties between the dE and dSph galaxies.

\noindent {\bf Figure 6.} Histograms of the distance of galaxies from the ZOE
defined by DHGs in the $\kappa_1$--$\kappa_2$ plane. The residuals are defined
as $\delta_{2:1} = \kappa_1 + \kappa_2 - 8$.  The diagram summarizes and
quantifies the march of Hubble types away from the ZOE in the previous figures.

\noindent {\bf Figure 7} Histograms of the distance of galaxies from the DHG
fundamental plane.  The residuals are defined as $\delta_{3:1} = \kappa_3 -
0.15\kappa_3 - 0.36$. Bin size is 0.05 dex.   This diagram reflects
conclusions from Figures~2--5: Sa--Sc galaxies define a fundamental plane that
is basically parallel to that for DHGs but offset by -0.2 dex to lower
$\kappa_3$ values.  Hubble types Scd--Irr define yet a third fundamental plane
with a different tilt and showing negligible slope in $\kappa_1$--$\kappa_3$.

\noindent {\bf Figure 8} The B-band Tully-Fisher relation for the spiral and
irregular galaxies in our sample. Symbols for the different Hubble types are
as in Figures 3--5.  A line of slope --7.5 (which corresponds to $A_C = 3$,
Equation~26) is drawn through the data. A line with $A_C = 4$ corresponds to a
plane that projects with minimal scatter onto $\kappa_2$-$\kappa_3$. 
Comparison of points here with $\kappa_2$-$\kappa_3$ in Figures~3--5 shows
systematic residuals that reflect $A_C = 3$ rather than $A_C = 4$.

\noindent {\bf Figure 9} $\kappa$ parameters for galaxy groups and galaxy
clusters.  Galaxy groups are divided into S-rich (triangles) and E-rich
(circles); within each class, groups are separated into those with 10 or more
members (closed symbols) or 9 or less (open symbols).  Data for 16 rich
clusters from Schaeffer et al. are plotted as large closed circles, and data
for two very large spiral galaxies Malin 1 and Malin 2 are plotted as large
crosses.  This 3D diagram samples a different region of $\kappa$-space than do
Figures~2--5 to accommodate the larger masses, lower surface brightnesses and
larger mass-to-light ratios for galaxy groups and clusters.

\noindent {\bf Figure~10} Histograms of the distance of galaxy groups and
galaxy clusters from the DHG fundamental plane. The residuals are the same as
in Figure~7, defined as $\delta_{3:1} = \kappa_3 - 0.15\kappa_3 - 0.36$.
Globular clusters and DHGs are included for reference.  For groups and
clusters, mean offset $\delta_{3:1}$ is correlated with both group size and
the mean Hubble type of member galaxies, and the distributions are wider for
poor groups of all types.  These behaviors are consistent with the strong
trend of $\kappa_3$ vs. $\kappa_1$ ($M/L$ vs $M$) in Figure~9.

\noindent {\bf Figure~11} The cosmic metaplane.  The $\kappa$ parameters for
all of the stellar systems studied in this paper.  Symbols for each kind of
objects are as in the previous $\kappa$ diagrams, with the addition of the
data for globular clusters, plotted as large closed triangles. At the large
scale of this 3D fold-out it is difficult to make out the separate Hubble
types for galaxies.  Yet, at this scale we can most easily see the overall
relationships among the $\kappa$ parameters for all these stellar systems,
which we now define as the cosmic metaplane.

\noindent {\bf Figure~12} (a) The distribution of all galaxies in our sample
(small solid squares) and the distribution of galaxy groups and clusters (open
and closed circles) in $\kappa_1$--$\kappa_2$.  $V_1$, $V_2$, and $V_3$  are
the three vectors connecting individual E galaxies with E-rich clusters
discussed in the text.  $V_1$ undoes the effects of baryonic dissipative
infall within dark halos (same as the separate vector labeled ``dissipation").
$V_2$ connects the dark halos of galaxies with those of E-rich groups along
the locus of an $n = -1.8$ power-law density fluctuation spectrum.  $V_3$ is a
final correction to the surface brightnesses of clusters owing to their higher
hot gas content.  Details of all vectors are given in the text.  The sum of
the three vectors is a fair match to the slope of the ZOE,
consistent with the combined predictions of hierarchical clustering/merging
and dissipation. (b) A schematic attempt to undo the effects of baryonic
dissipation in galaxy halos by sliding galaxies back along the dissipation
vector of Figure~12a. The shifted positions show galaxies as they would appear
if their stars were redistributed with the same spatial distribution as their
dark matter.  The recovered, pre-dissipation halos blend smoothly with groups
and clusters, showing that the previous separation was due to dissipation. 
For this shift, a shrinkage in $M_e$ and $R_e$ by a factor of 10 was assumed,
with no change in $\sigma_c$.

\noindent {\bf Figure 13} The analogue to Figure~3 in BFPR, plotting baryonic
number density versus effective kinetic temperature for all stellar systems. 
Conversion formulae from $\kappa$-parameters to $n_{bary,e}$ and $T_e$ are
given in the text.  The distribution of galaxies here using real data is fat
and horizontal, quite different from the thin, tilted distribution using
schematic data in BFPR.  The slope of clustering from a power-law density
fluctuation spectrum with constant dissipation and $n = -1.8$ is shown.

\noindent {\bf Figure 14} The analogue to Figure~4 in
BFPR, in which we now plot $M_e$ (in units of solar masses) versus 
effective kinetic temperature.  The slope of a power-law density
fluctuation spectrum with constant dissipation and $n = -1.8$ is shown.

\noindent {\bf Figure 15} The companion diagram to Figures~13 and 14, plotting
a direct analogue to the Tully-Fisher relationship using mass and radius.  The
slope of a power-law density fluctuation spectrum with constant dissipation
and $n = -1.8$ is shown.

\noindent {\bf Figure 16} (a) The predicted distribution within $\kappa$-space
for the CHDM$_2$ model of Klypin et al. (1993), based on groupings made in
redshift space.  This simulation mimics the manner in which groups are
constructed using real data.  The distribution of these mock-groups is a
reasonable match to that of real groups. (b) The predicted distribution within
$\kappa$-space for the CHDM$_2$ model of Klypin et al. (1993), based on
groupings made in distance space (i.e., using fully three dimensional
information that is not generally available with real data).  It is reassuring
that the distribution of these distance-selected mock-groups is quite similar
to those both for real groups and for mock-groups selected in redshift space.

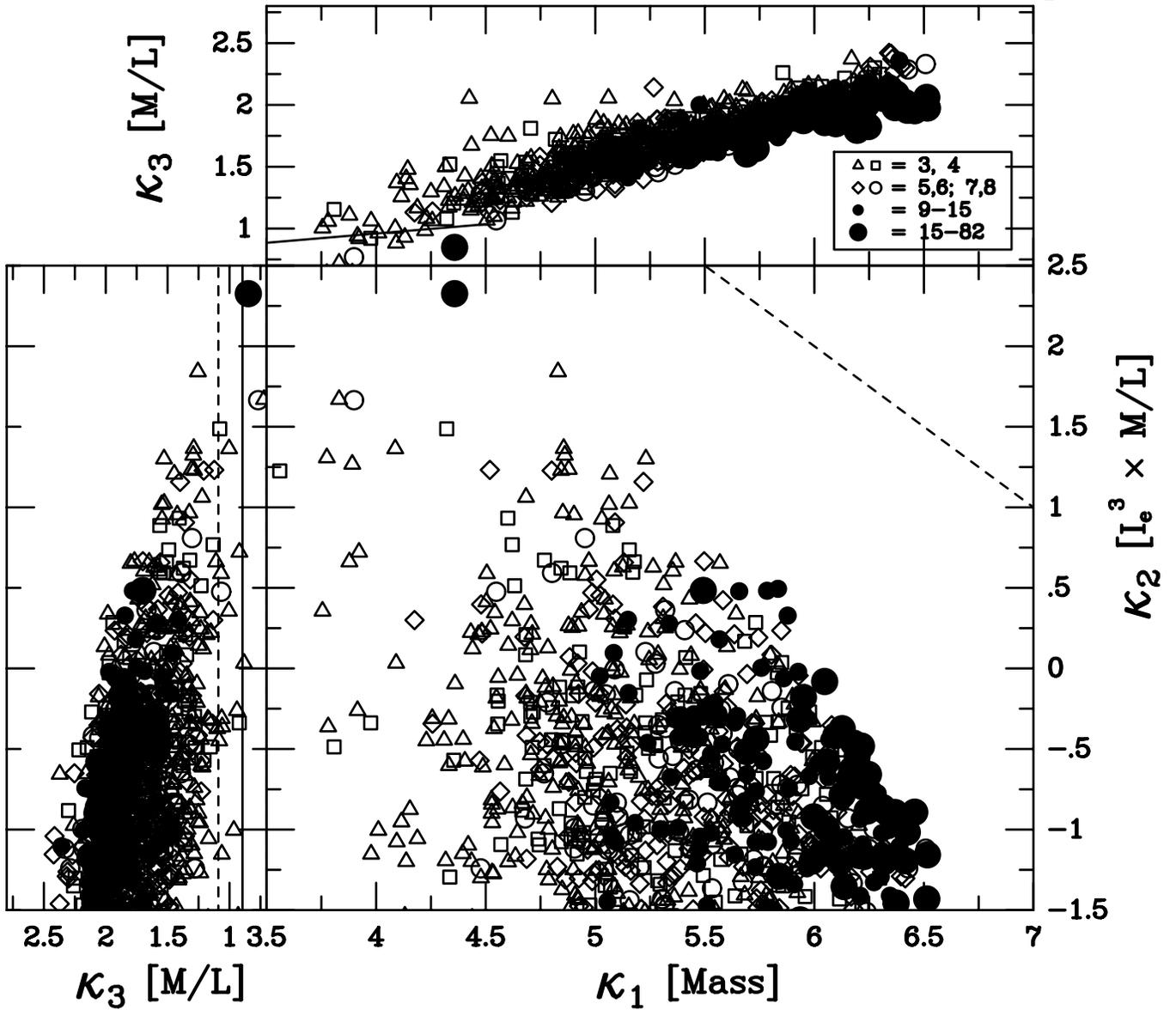
\end{document}